\documentclass[11pt]{article}
\usepackage{graphics,epsfig}
\usepackage[latin1]{inputenc}
\usepackage[english,activeacute]{babel}
\usepackage[]{graphicx}
\usepackage{latexsym}
\usepackage{amsmath, amsthm, amsfonts, amssymb}
\usepackage{verbatim}

\begin{document}

\newtheorem{theo}{Theorem}[section]
\newtheorem{definition}[theo]{Definition}
\newtheorem{lem}[theo]{Lemma}
\newtheorem{prop}[theo]{Proposition}
\newtheorem{coro}[theo]{Corollary}
\newtheorem{exam}[theo]{Example}
\newtheorem{rema}[theo]{Remark}
\newtheorem{example}[theo]{Example}
\newtheorem{principle}[theo]{Principle}
\newcommand{\ninv}{\mathord{\sim}}
\newtheorem{axiom}[theo]{Axiom}

\title{Towards a definition of the Quantum Ergodic Hierarchy: Kolmogorov and
Bernoulli systems}
\author{\textsc{Mario Castagnino}$^{1}$ \ \textsc{and} \ \textsc{Ignacio Gomez}$%
^{2}$}
\maketitle

\begin{abstract}
In this paper we translate the two higher levels of the Ergodic Hierarchy \cite{3M},
the Kolmogorov level and the Bernoulli level, to quantum language. Moreover, this
paper can be considered as the second part of \cite{0}. As
in paper \cite{0}, we consider the formalism where the states are positive
functionals on the algebra of observables and we use the properties of the
Wigner transform \cite{...W}. We illustrate the physical relevance of the
Quantum Ergodic Hierarchy with two emblematic examples of the literature:
the Casati-Prosen model \cite{casati verdadero}, \cite{casati model} and the kicked rotator \cite{stockmann}, \cite{casati libro}, \cite{haake}.
\end{abstract}

\begin{center}
{\small 1- Instituto de Física de Rosario (IFIR-CONICET) and \\[0pt]
Instituto de Astronomía y Física del Espacio, \\[0pt]
Casilla de Correos 67, Sucursal 28, 1428 Buenos Aires, Argentina.\\[0pt]
2- Instituto de Física de Rosario (IFIR-CONICET), Rosario, Argentina\\[0pt]
}
\end{center}

\vspace{1cm}

\bigskip \noindent

{\small \noindent
\centerline{\emph{Key words:
Ergodic-Mixing-Kolmogorov-Bernoulli-EH-QEH}} }

\section{Introduction}

This paper is based on \cite{bellot} which presents a three theories
structure of classical chaos: Ergodic Hierarchy, Lyapunov exponents and Complexity. The first
two theories are related to the Pesin's theorem \cite{lich} and
the last two ones to the Brudno's theorem \cite{bellot}. The Pesin's theorem
expresses the equivalence between the KS entropy and the exponential
divergence of trajectories by the presence of positive Lyapunov exponents,
and the positivity of these exponents is a necessary and sufficient
condition for chaos. On the other hand, the Brudno's theorem expresses the
equivalence between the complexity of almost every point of the phase space and the KS entropy. The theoretical relation
between these chaos indicators, KS entropy, complexity and Lyapunov
exponents are sketched in a ``chaos pyramid" in the figure 1. According to
this structure Ergodic Hierarchy is one of the features of classical chaos.
We have presented in our previous paper \cite{0} a proposal to define the
first two steps of a Quantum Ergodic Hierarchy, i.e. Quantum Ergodic and
Quantum Mixing systems. Then the only purpose of this paper is to complete
the Quantum Ergodic Hierarchy with two more steps: Quantum Kolmogorov and Quantum Bernouilli.
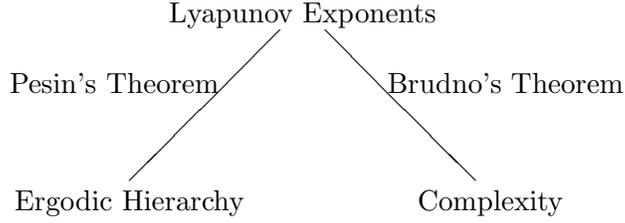
\begin{figure}[tbp]
\label{pyramid}
\par
\begin{center}
\unitlength=1mm
\begin{picture}(5,23)(0,0)
\put(-3,23){\line(-1,-1){20}} \put(3,23){\line(1,-1){20}}
\put(0,25){\makebox(0,0){Lyapunov Exponents}}
\put(-23,0){\makebox(0,0){Ergodic Hierarchy}}
\put(25,0){\makebox(0,0){Complexity}}
\put(27,16){\makebox(0,0){Brudno's Theorem}}
\put(-25,16){\makebox(0,0){Pesin's Theorem}}
\end{picture}
\end{center}
\caption{The ``chaos pyramid" is a diagram of the relationships beetween the
three theories structures: Ergodic Hierarchy, Lyapunov and Complexity
through the Pesin and Brudno theorems.}
\end{figure}
On the other hand the large majority of works on Quantum Chaos follows a
different line. That is, in books \cite{gutzwiller}, \cite{benati}, \cite{stockmann}, \cite{casati libro},
and \cite{haake} only
``bra" and ``kets" appear. Concepts like ``observables" and ``functionals" are not taken into account in a fundamental way in these books. Then a crucial
definition of the mixing systems does not appear in these books, i.e. as the
property of a quantum system with weak limit reaches equilibrium \cite{0}.
Essentially the way to introduce a Quantum Ergodic Hierarchy cannot be based
in the study of quantum closed systems with just ``bra" and ``kets". Other
concepts must be introduced as ``observables" and ``mixed states".

More precisely, following the Ballentine's book \cite{Ballentine}, where
an axiomatic structure for Quantum Mechanics is sketched, the primitive
concepts of Observable $O$ are introduced. Then the states $\rho$ are defined
as a derived concept. They are the functional over the observable space.
Then the ``bra" and ``kets" are simple vectors while the observables and
states are matrices. Then somehow the usual treatment of quantum chaos in
closed systems with the ``bra" and ``ket" is not enough, since this
formalism does not consider open systems.

Moreover, as we explained in the introduction of \cite{0}, there
are many ways to define quantum chaos. Among these ways,
following part one, we will say that \emph{``a quantum system is chaotic if its
classical limit is chaotic"}, i.e. the Michael Berry's quantum chaos definition \cite{BERRY}. In
this way we have defined quantum chaos in the two steps of the Quantum
Ergodic Hierarchy: Ergodic and Mixing \cite{0}. These steps are defined by
their correspondent classical limit: they have a Cèsaro and a Weak limit
respectively, that correspond to the limit of the Ergodic and Mixing systems of the classical ergodic hierarchy \cite{3M}. We will follow the same
strategy with the other two levels.

Then, we have explained this limit in great detail. Precisely we have considered
the Weyl-Wigner-Moyal transformation \cite{...W}, the
transformation $symb$ that changes quantum operators and states into the
corresponding classical symbols which become operators and states of classical analytic Mechanics when $\hbar \rightarrow 0$.
\begin{equation*}
symb\widehat{O}=O(q,p),\text{ \ \ }symb\widehat{\rho }=\rho (q,p),\text{ \
if }\hbar \rightarrow 0
\end{equation*}%
Then it can be demonstrated that
\begin{equation}
Tr(\widehat{\rho }\widehat{O})=\int_{\Gamma }\rho (q,p)O(q,p)dqdp  \label{Eq}
\end{equation}%
where $\Gamma $ is the phase space, and the last equation is valid even if
$\hbar \neq 0$.

Thus, we define a classical mixing system as the one that satisfies the
weak limit
\begin{equation*}
\lim_{t\rightarrow \infty }\int_{\Gamma }\rho (q,p,t)O(q,p)dqdp=\int_{\Gamma
}\rho _{\ast }(q,p)O(q,p)dqdp
\end{equation*}%
where $\rho _{\ast }(q,p)$ is the weak equilibrium state of $\rho(q,p,t)$ and $O(q,p)$ belongs to the space of observables. Then from \eqref{Eq} the corresponding quantum chaos
will satisfy the weak limit
\begin{equation*}
\lim_{t\rightarrow \infty }Tr(\widehat{\rho }(t)\widehat{O})=Tr(\widehat{%
\rho }_{\ast }\widehat{O})
\end{equation*}%
namely the definition of quantum mixing chaos that we have given in paper
\cite{0} section 6.3, definition B.

This way to define the quantum version of the Ergodic Hierarchy, based in
the classical one, is the one that we will also use in this paper (see
Table II page 28).

As a consequence, following the ideas of \cite{bellot} and the preceding paper \cite{0}, we will study the problem of quantum chaos
hierarchy \textit{directly from the quantum description of the chaotic classical limit}.

So as in paper \cite{0}, that we consider as the\textit{\ first part} of
this paper, we have defined the quantum chaos in the two first levels of the
ergodic hierarchy (EH): ergodic and mixing. In this paper we will complete
the work adding two more levels: Kolmogorov and Bernoulli. Then this paper
can be considered as a \textit{second part} of paper \cite{0}. Nevertheless
in this paper following the ideas of paper \cite{3M} we will first repeat
the two initial levels using these new concepts and then adding the two final
levels. Also, for the sake of conciseness, we will not repeat the following
sections of paper \cite{0}: section 2 (Mathematical background), section 3
(Decoherence in non integral systems), section 4 (The classical statistical
limit), and section 5 (The classical limit). These sections can be read in
\cite{0}. The just quoted section 5 can also be complemented with paper \cite{FOP}.

The paper is thus organized. Section 2: We present the formalism,
definitions and the Ergodic Hierarchy (EH) which we will use. Section 3
and 4: we briefly review the ergodic and mixing systems already considered
in ref. \cite{0}. Section 5 and 6: We explain the Kolmogorov and
Bernoulli cases in detail. Section 7: we give a physical relevance analyzing two emblematic examples of the literature in terms of the Quantum Ergodic
Hierarchy: the Casati-Prosen model \cite{casati verdadero, casati model} and the kicked rotator \cite{stockmann, casati libro, haake}. Section 8: We consider the relevance of the subject and draw our
conclusions.

\section{Formalism}

\label{s:preliminaries}

\noindent For a Hilbert space $\mathcal{H}$ of dimension $N > 2$ the set of
pure states forms a $(2N - 2)$-dimensional manifold, of measure zero, in the
$(N^2 - 2)-$dimensional boundary $\partial \mathcal{C}_N$ of the set $%
\mathcal{C}_N$ of density matrices. The set of mixed quantum states $%
\mathcal{C}_N$ consists of Hermitian, positive matrices of size $N$,
normalized by the trace condition, that is

\begin{eqnarray}
\mathcal{C}_N= \{\rho: \rho = \rho^{\dagger};\,\,\, \rho \ge 0;\,\,\,
tr(\rho) = 1;\,\,\, dim(\rho) = N\}.
\end{eqnarray}

\noindent It can be shown for finite dimensional bipartite states that there
exists always a non-zero measure $\mu_s$ in the neighborhood of separable
states containing maximum uncertainty ones $\mu_s$ tends to zero as the
dimension tends to infinity. Finally, for an infinitely dimensional Hilbert
space almost all states are entangled \cite{CHK1,CHK2}.

\noindent In this section we would like to establish the relation of three
different levels:

\begin{itemize}
\item The notion of \textit{sets} \textit{correlations}, the main tool of
paper \cite{3M}.

\item The \textit{algebra} of the observables and states (symbolized by
density or distribution functions) in the phase space of the Hamiltonian
Mechanics (see \cite{LM} and \cite{M}).

\item The same algebra at \textit{quantum mechanics level} obtained through
the Weyl-Wigner-Moyal transformation form the Hamiltonian Mechanics (see
\cite{...W} and \cite{Dito} for details).
\end{itemize}

\subsection{Definitions}

\begin{itemize}
\item Let $[X,\Sigma ,\mu ,T_{t}]$ be a generic \textit{dynamically system, }%
where $X$ is a set , $\Sigma $ is a $\sigma -$algebra, $\mu $ is a measure,
and $T_{t}$ is a time transformation. In Hamiltonian Mechanics, the physical
case, $X$ is the phase space with coordinates $\phi =(q,p)$, (or a
projection $\Pi X$ of phase space whose coordinates will also symbolize $%
\phi $), $\Sigma $ is the $\sigma -$algebra of measurable sets of $X$, $\mu $
is the Liouville measure $d\phi =dqdp$, (usually normalized as $\mu (X)=1)$,
and $T_{t}$ is given by the Hamiltonian dynamics.

\item The $\Sigma $ algebra has the following properties:

(1) $X \in \Sigma ,$

(2) $A\backslash B \in \Sigma $ for all $A,B \in \Sigma ,$ and

(3) $\bigcup_{i=1}^{n}B_{i} \in \Sigma $ if $B_{i} \in \Sigma $ for $1\leq
i\leq n\leq \infty .$ As a consequence $\Sigma $ also contains $\varnothing $
and $\bigcap_{i=1}^{n}B_{i}$ if $B_{i} \in \Sigma ,$ for $1\leq i\leq n\leq
\infty $

\item The probability measure $\mu $ on $\Sigma $ is such that

(1) $\mu :\Sigma \rightarrow 1$ with $\mu (X)=1,$ and

(2) If $\{B_{i}\}_{i=1}^{n}\subset \Sigma $ and $B_{j}\cap B_{k}=\varnothing
$ for $1\leq j\leq k\leq n\leq $ then $\mu \left(
\bigcup_{i=1}^{n}B_{i}\right) =\sum_{i=1}^{n}\mu (B_{i}).$

\item The automorphism $T$ is an automorphism that maps the probability
space $[X,B,\mu ]$ onto itself \ and it is measure preserving iff $B \in X$
i.e.:

(1) $T^{-1}B \in \Sigma ,$

(2) $\mu (T^{-1}B)=\mu (B),$where $T^{-1}B=[x \in X:Tx \in B]$

\item A dynamic law or time evolution $\tau =\{T_{t}\}_{t\in I}$ is a group
of measure preserving authomorphisms $T_{t}X\rightarrow X$\ of the
probability space $[X,B,\mu ]$ onto itself and where $I$ is either $\mathbb{R%
}$ or $\mathbb{Z.}$

\item The set $\alpha =\{\alpha _{i}:i=1,...,N)$ is a partition of $X$ iff

(1) $\alpha _{i}\cap \alpha _{j}=\varnothing $ for all $i\neq j,$

(2) $\mu (X\backslash \bigcup_{i=1}^{n})=0.$

(3) Given two partitions $\alpha =\{\alpha _{i}:i=1,...,N\},$ $\beta
=\{\beta _{j}:j=1,...,M\}$ we will call their sum $\alpha \vee \beta
=\{\alpha _{i}\cap \beta _{j}:i=1,...,N;j=1,...,M\}$

\item A $\sigma -$sub algebra $\Sigma _{0}\subset \Sigma $ must also satisfy
the conditions

(1) $\Sigma _{0}\subseteq T\Sigma _{0},$

(2) $\vee _{n=-\infty }^{\infty }T^{n}\Sigma _{0}=\Sigma ,$

(3) $\wedge _{n=-\infty }^{\infty }T^{n}\Sigma _{0}=N$ namely the $\sigma -$%
algebra containing the set of measure one and zero\footnote{%
Of course, the last condition cannot be fulfilled in the quantum case
because the phase space will have an intrinsic graininess originated in the
uncertainty principle. So we must consider that this condition would only be
approximately satisfied.}.

\item Let $A$ and $B$ be measurable sets of the space $X$, and let $\mu $ be
the measure just defined. Then the \textit{correlation }between $A$ and $B $
is defined as%
\begin{equation}
C(B,A)=\mu (A\cap B)-\mu (A)\mu (B)  \label{2.0}
\end{equation}%
Let us explain the meaning of this notion. In a generic system and under
generic circumstances we have $C(B,A)\neq 0.$ But if $C(B,A)=0$ some kind of
\textquotedblleft homogeneity" has appeared in the system since both factors
$\mu (A)$ and $\mu (B)$ play the same role in the product $\mu (A\cap B)$,
precisely:

\begin{equation}
\mu (A\cap B)=\mu (A)\mu (B)  \label{H}
\end{equation}%
This \textquotedblleft homogenization" in the behavior of $\mu (A\cap B)$
corresponds to the vanishing of correlations. Then, if the time evolution $%
T_{t}$ conserves the measures or $\mu (T_{t}A)=\mu (A),$ (as in the phase
space) we have

\begin{equation}
\mu (T_{t}A\cap B)=\mu (A)\mu (B)+C(B,T_{t}A)  \label{2}
\end{equation}

where $\mu (A)\mu (B)$ would be the \textquotedblleft homogenous" constant
part of $\mu (T_{t}A\cap B)$ and $C(B,T_{t}A)$ the \textquotedblleft
non-homogenous" variable part. Then, if e. g., when $t\rightarrow \infty $
we have $C(B,A)\rightarrow 0$ some homogenization has taken place in the
system \footnote{%
Correlations are also related with the notion of \textit{unpredictability}
\cite{3M}, but we do not consider this subject in this paper.}.
\end{itemize}

\subsection{The Ergodic Hierarchy (EH)}

Using the notion of \textit{correlation} (see equations \eqref{2.0} and %
\eqref{2}) we can define the main four steps of the EH as:

\begin{itemize}
\item \textit{Ergodic} systems if
\begin{equation}
\lim_{T\rightarrow \infty }\frac{1}{T}\int_{0}^{T}C(T_{t}B,A)dt=0  \label{ER}
\end{equation}

We will call this limit a \textit{Cesàro} limit.

\item \textit{Mixing} systems if

\begin{equation}
\lim_{t\rightarrow \infty }C(T_{t}B,A)=0  \label{MI}
\end{equation}

We will call this limit a \textit{weak} limit.

\item \textit{Kolmogorov }systems. Using the Cornfeld-Fomin-Sinai
theorem (see \cite{CFS} page 283) the traditional definition for these system
can be translated to the correlations´ language as follows:

\textit{A system is Kolmogorov (K-system) if for any integer }$r$\textit{\
and any set }$A_{0},A_{1},A_{2},..A_{r} \in X$\textit{\ and for any }$%
\varepsilon >0$\textit{\ there exists an }$n_{0}>0$\textit{\ such for all }$%
B \in \sigma _{n,r}(A_{1},A_{2},..A_{r}),$\textit{\ we have}

\begin{equation}
|C(B,A_{0})|<\varepsilon  \label{KO}
\end{equation}

\textit{where }$\sigma _{n,r}(A_{1},A_{2},..A_{r})$\textit{\ is a sub }$%
\sigma -$\textit{algebra} (defined in section 2.1.vii)

For example this $\sigma -$algebra contains, among others, the following
sets.

\begin{enumerate}
\item All the $T_{k}A_{i},$ for all $k\geq n,$ and all $i=1,...r.$

\item All the finite and infinite sequences $T_{n}A_{m_{1}}\cup
T_{n}A_{m_{2}}\cup T_{n}A_{m_{3}}...,$ and $T_{n}A_{m_{1}}\cup
T_{n+1}A_{m_{2}}\cup T_{n+2}A_{m_{3}}...$ where $m_{i} \in (i=1,...r)$

\item All the finite and infinite sequences $T_{n} A_{m_{1}} \cap
T_{n}A_{m_{2}}\cap T_{n}A_{m_{3}}...$ and $T_{n}A_{m_{1}}\cap
T_{n+1}A_{m_{2}}\cap T_{n+2}A_{m_{3\text{ }}}...\frac{{}}{{}}$where $m_{i}
\in (i=1,...r)$
\end{enumerate}

\item \textit{Bernoulli} system If for any time $t$

\begin{equation}
C(T_{t}B,A)=0  \label{BE}
\end{equation}

\noindent so from eq. (\ref{2}) if $\mu (T_{t}A)=\mu (A)$ we have

\begin{equation}
\mu (T_{t}B\cap A)=\mu (A)\mu (B)  \label{7'}
\end{equation}

i.e. in probability language: the probability to obtain the event $B,$ at
any time, conditioned by $A$ is always the same and we have the homogeneity
defined in eq. (\ref{H}).

Then the levels of the EH are defined \textit{by the way the correlations
vanishes} when $t\rightarrow \infty $ (being the Bernoulli level defined by
a trivial zero identity).
\end{itemize}

\subsection{Correlations at the different levels}

Now we can also define the notion of correlation at their different levels
of subsection 2.2.

\emph{I}) \textit{Measurable set} level
\begin{equation}
C(B,A)=\mu (A\cap B)-\mu (A)\mu (B)  \label{2.1'}
\end{equation}

\emph{II}) \textit{Distribution or density function} level

\begin{equation}
C(g,f)=\langle f,g\rangle -\langle f,1\rangle \langle 1,g\rangle
\label{2.2''}
\end{equation}%
where $f$ (and $g)$ is a function over the phase space $X$ such that the
integral $\int_{X}f(\phi )d\phi $ exists, $\langle f,g\rangle
=\int_{X}f(\phi )g(\phi )d\phi $ and where $\phi =(q,p),$ are the
coordinates at a point of $X,$ so $\phi \in X$ and $d\phi =\mu (d\phi
)=dqdp. $

\emph{III}) \textit{Quantum} level
\begin{equation}
C(\widehat{g}|\widehat{,f})=(\widehat{f}|\widehat{g})-(\widehat{f}|\widehat{I%
})(\widehat{I}|\widehat{g})  \label{2.3'}
\end{equation}%
where $\widehat{f},\widehat{g}\in \mathcal{A}$ the algebra of
observables. Then if $f=symb(\widehat{f}|$ and $g=symb(\widehat{g}|$ are the
Weyl-Wigner-Moyal transforms of $(\widehat{f}|$ and $|\widehat{g})$ we know
that $(\widehat{f}|\widehat{g})=\langle f,g\rangle $\footnote{%
In the process, from \emph{I}) to \emph{III}), we may say that the ignorance
probabilities become intrinsic probabilities, but numerically they are equal.%
}. Then using the usual quantum symbols for observables and states we
have%
\begin{equation}
C(\widehat{O},\widehat{\rho })=(\widehat{\rho }|\widehat{O})-(\widehat{\rho }%
|\widehat{I})(\widehat{I}|\widehat{O})  \label{2.3''}
\end{equation}%
where $\widehat{\rho }=\widehat{f}$ are the states and $\widehat{O}=\widehat{%
g}$ are the observables.\footnote{%
The normalization of $\widehat{\rho }(t)$ is simply $(\widehat{\rho }|%
\widehat{I})=1$ or $Tr\widehat{\rho }=1$ so $C(\widehat{O},\widehat{\rho })=(%
\widehat{\rho }|\widehat{O})-(\widehat{I}|\widehat{O})=(\widehat{\rho }|%
\widehat{O})-Tr\widehat{O}$}

From these equations we can see that we can translate the EH up to a
Quantum Ergodic Hierarchy (QEH), we have done for the two first
steps, for The Ergodic Hierarchy, in paper \cite{0}.

Let us now schematically show the relations among eqs., (\ref{2.1'}) to (\ref%
{2.3''}). Let us define the characteristic function $1_{A}(\phi )$ as
\begin{equation*}
1_{A}(\phi )=1\text{ if }\phi \in A,\text{ \ \ \ }1_{A}(\phi )=0\text{ if }%
\phi \notin A
\end{equation*}%
Then as $1_{A}^{2}(\phi )=1_{A}(\phi ),$ and $1_{A}(\phi )$ can also be
considered as a projector $\Pi _{A}(\phi )=1_{A}(\phi ).$ Using these
projectors we can write the definition (\ref{2.1'}) as%
\begin{equation}
C(B,A)=\int_{X}1_{A}(\phi )1_{B}(\phi )d\phi -\int_{X}1_{A}(\phi )d\phi
\int_{X}1_{B}(\phi )d\phi  \label{2.2}
\end{equation}%
since it is evident that the terms of the r.h.s. \ of both equations are the
same.

Let us now define a partition $\{A_{i}\}$ of $X$ that satisfies%
\begin{equation*}
X=\bigcup_{i}A_{i},\text{ \ \ \ \ }A_{i}\cap A_{i}=\emptyset \text{ \ if }%
i\neq j
\end{equation*}%
or such that
\begin{equation*}
1_{A_{i}}1_{A_{j}}=\delta _{ij}1_{A_{i}}
\end{equation*}%
Let us also introduce two arbitrary sets of number $a_{i},b_{j}\epsilon
\mathbb{R},$ then from eq. (\ref{2.2})%
\begin{equation*}
\sum_{ij}a_{i},b_{j}C(A_{j},A_{i})=
\end{equation*}%
\begin{equation}
=\sum_{ij}\int_{X}a_{i},b_{j}1_{A_{i}}(\phi )1_{A_{j}}(\phi )d\phi
-\sum_{i}\int_{X}a_{i}1_{A_{i}}(\phi )d\phi
\sum_{j}\int_{X}b_{j}1_{A_{j}}(\phi )d\phi  \label{2.2'}
\end{equation}%
Then if we define two functions
\begin{equation*}
f(\phi )=\sum_{i}a_{i}1_{A_{i}}(\phi ),\text{ \ \ \ \ }g((\phi
)=\sum_{j}b_{j}1_{A_{j}}(\phi )
\end{equation*}%
\noindent it is clear that since we can make the domains $A_{i}$ of the
partition as small as we want we can approximate $f(\phi)$ and $g(\phi)$, then we can define
\begin{equation*}
C(g,f)=\sum_{ij}a_{i},b_{j}C(A_{j},A_{i})=\int_{X}f(\phi )g(\phi )d\phi
-\int_{X}f(\phi )d\phi \int_{X}g(\phi )d\phi
\end{equation*}%
\noindent or defining $\langle f(\phi ),g(\phi )\rangle =\int_{X}f(\phi
)g(\phi )d\phi .$
\begin{equation}
C(g,f)=\sum_{ij}a_{i},b_{j}C(A_{j},A_{i})=\langle f(\phi ),g(\phi )\rangle
-\langle f(\phi ),1\rangle \langle 1,g(\phi )\rangle  \label{2.8}
\end{equation}

\noindent i.e. the definition of correlations in the \textit{%
distribution function language} (cf. eq. (\ref{2.2''}))\ is demonstrated.
This definition is equivalent to (\ref{2.1'}) if $a_{i}=\delta
_{i0},b_{j}=\delta _{j1},A_{0}=B,A_{1}=A$.

Given $(\widehat{\rho }|\widehat{I})=\langle \rho ,1\rangle =\langle
\rho \rangle$ and $symb\widehat{I}=1$, using to \eqref{2.8} the
Weyl-Wigner-Moyal transform and interpreting $\widehat{f}$ as the state and $%
\widehat{g}$ as the operator, if $symb\widehat{O}=O(\phi )$ and $symb%
\widehat{\rho }=\rho (\phi )$, we have that

\begin{equation}
C(\widehat{O},\widehat{\rho })=(\widehat{\rho }|\widehat{O})-(\widehat{\rho }%
|\widehat{I})(\widehat{I}|\widehat{O})  \label{2.9}
\end{equation}

\noindent i.e. the definition of correlations but now at the \textit{quantum
language }(cf. eq. (\ref{2.3''})) which is equivalent to (\ref{2.8}) from
the properties of Weyl-Wigner-Moyal transform. So when $\hbar \rightarrow 0$
we have (\ref{2.9})$\Leftrightarrow $(\ref{2.8})$\Leftrightarrow $(\ref{2.1'}%
).

So we can see that the three levels: measurable set level, distribution
function level, and quantum level are all equivalent and interchangeable.

\subsection{More general equations and the Ergodic Hierarchy (EH)}

\begin{itemize}
\item We will call Frobenius-Perron operator $P_{t}$ to the evolution
operator of distributions or density functions. In quantum language the
Frobenius-Perron operator $P_{t}$ would be the evolution operator for
states, while the Koopman operator $U_{t}$ would be the time evolution
operator for observables. In fact we have that%
\begin{equation}
\langle P_{t}f,g\rangle =\langle f,U_{t}g\rangle  \label{FPK}
\end{equation}%
see \cite{LM} eq.(3.3.4).

Then $P_{t}$, the Frobenius-Perron operator, conserves the measure. Then we
have$\int_{X}P_{t}1_{A_{i}}d\phi =\int_{X}1_{A_{i}}d\phi $ and $%
\sum_{i}a_{i}\int_{X}P_{t}1_{A_{i}}d\phi =\sum_{i}a_{i}\int_{X}1_{A_{i}}$
thus%
\begin{equation}
\int_{X}P_{t}fd\phi =\int_{X}fd\phi \text{ or }\langle P_{t}f\rangle
=\langle f\rangle  \label{C'}
\end{equation}%
\noindent or at the quantum level, since $\langle f\rangle =\langle
f.I\rangle =(\widehat{f|}\widehat{I})=Tr\widehat{f}$, we have
\begin{equation}
Tr(\widehat{\rho }(t))=Tr(\widehat{\rho }(0))  \label{C}
\end{equation}%
\noindent namely the trace is also conserved.

\item In general there exists several $f_{\ast }$, the \textit{%
equilibrium distributions} such, that $P_{t}f_{\ast }=f_{\ast }$. But if the
system is ergodic there is only one of them, therefore we will only consider
this case.

\item At the two first levels of the EH we will have a limit (Cesàro,
Mixing) $P_{t}f\rightarrow f_{\ast }$ when $t\rightarrow \infty $ and from
this limit we will have $\langle f_{\ast }\rangle =\langle f(t)\rangle $ or $%
Tr\widehat{\rho }_{\ast }=Tr\rho (t)$, since the norm is also conserved at
the limit.

Then we can define a \textit{new measure} $\mu _{\ast }(A)$ such that%
\begin{equation*}
\mu _{\ast }(A)=\int_{A}f_{\ast }(\phi )d\phi
\end{equation*}%
and define a new correlation%
\begin{equation*}
C_{\ast }(B,A)=\mu _{\ast }(A\cap B)-\mu _{\ast }(A)\mu _{\ast }(B)
\end{equation*}%
Now we can define the new levels: \textit{Ergodic and Mixing }making $\mu
\rightarrow \mu _{\ast }$ in eqs. (\ref{ER}) to (\ref{BE}). So we have the Ergodic Hierarchy (EH). Then, e. g., in the mixing case (see \cite%
{Mackeylibro} pag. 58) we have
\begin{equation}
\lim_{t\rightarrow \infty }\mu _{\ast }(T_{t}A\cap B)=\mu _{\ast }(A)\mu
_{\ast }(B)  \label{2.1}
\end{equation}%
\noindent then
\begin{equation*}
\lim_{t\rightarrow \infty }\mu _{\ast }(T_{t}A\cap X)=\mu _{\ast }(A)\mu
_{\ast }(X)
\end{equation*}%
\noindent and if we normalize $\mu _{\ast }(X)=1.$%
\begin{equation*}
\lim_{t\rightarrow \infty }\mu _{\ast }(T_{t}A)=\mu _{\ast }(A)
\end{equation*}%
\noindent i.e. the conservation of the normalization is also valid at the
limit $t\rightarrow \infty$.

\item Let us quote the Theorem 5.1 of \cite{Mackeylibro}:

\textquotedblleft Let $T_{t}$ be an ergodic transformation, with stationary
density $f_{\ast }(\phi )$ of the associated Frobenius-Perron operator,
operating in a phase space of finite $\mu _{\ast }$ measure. Then $T_{t}$ is
mixing iff $\{P_{t}f\}$ is weakly convergent to $f_{\ast }(\phi )$ for all
densities $f$, i. e.
\begin{equation*}
\lim_{t\rightarrow \infty }\langle P_{t}f,g\rangle =\langle f_{\ast
},g\rangle
\end{equation*}%
\noindent for every bounded measurable function $g$".

The demonstration is:
\begin{equation}
\begin{split}
& \lim_{t\rightarrow \infty }\mu _{\ast }(T_{t}A\cap B)=\lim_{t\rightarrow
\infty }\int_{T_{t}A\cap B}f_{\ast }(\phi )d\phi =\lim_{t\rightarrow \infty
}\int_{X}1_{T_{t}A\cap B}f_{\ast }(\phi )d\phi = \\
& =\lim_{t\rightarrow \infty }\int_{X}1_{T_{t}A}1_{B}f_{\ast }(\phi )d\phi
=\lim_{t\rightarrow \infty }\langle P_{t}1_{A}f_{\ast }(\phi ),1_{B}\rangle
\end{split}%
\end{equation}%
\noindent and also
\begin{equation*}
\mu _{\ast }(A)\mu _{\ast }(B)=\int_{X}1_{A}f_{\ast }(\phi )d\phi
\int_{X}1_{B}f_{\ast }(\phi )d\phi =\langle 1_{A}f_{\ast }(\phi ),1\rangle
\langle f_{\ast }(\phi ),1_{B}\rangle
\end{equation*}%
so from eq. (\ref{2.1}) we have%
\begin{equation*}
\lim_{t\rightarrow \infty }\langle P_{t}1_{A}f_{\ast }(\phi ),1_{B}\rangle
=\langle 1_{A}f_{\ast }(\phi ),1\rangle \langle f_{\ast }(\phi ),1_{B}\rangle
\end{equation*}%
or%
\begin{equation*}
\lim_{t\rightarrow \infty }\langle P_{t}1_{A_{i}}f_{\ast }(\phi
),1_{B_{j}}\rangle =\langle 1_{A_{i}}f_{\ast }(\phi ),1\rangle \langle
f_{\ast }(\phi ),1_{B_{j}}\rangle
\end{equation*}%
so considering two sets of generic numbers $(a_{i})$ and $(b_{j})$ and
define the generic functions
\begin{equation*}
f=\sum_{i}a_{i}1_{A_{i}}f_{\ast }(\phi ),\text{ \ \ \ }g=%
\sum_{j}b_{j}1_{B_{j}},\text{ }
\end{equation*}%
we obtain%
\begin{equation*}
\lim_{t\rightarrow \infty }\langle f,g\rangle =\langle f,1\rangle \langle
f_{\ast }(\phi ),g\rangle
\end{equation*}%
and if $f$ is normalized as $\langle f,1\rangle =1$ \ the thesis follows.
q.e.d.

Or in other words,
\begin{equation}
W-\lim_{t\rightarrow \infty }P_{t}f=f_{\ast }
\end{equation}
Finally the corresponding definition of quantum mixing is

\begin{equation}
\lim_{t\rightarrow \infty }(\widehat{\rho }(t)|\widehat{O})=(\widehat{\rho }%
_{\ast }|\widehat{O})  \label{2.11}
\end{equation}%
namely $\widehat{\rho }(t)$ weakly converges to $\widehat{\rho }_{\ast }$
(see \cite{0}).

For the \textit{ergodic} case we must simply make the substitution $%
\lim_{t\rightarrow \infty }\rightarrow \lim_{t\rightarrow \infty }\frac{1}{T}%
\int_{0}^{T}$ or $\lim_{n\rightarrow \infty }\frac{1}{n}\sum_{0}^{n-1}$ in
the discrete case. The Kolmogorov and Bernoulli cases will be considered in
sections 5 and 6. In Table I we display the synthetic structure of the three levels. \vskip1truecm
\centerline{TABLE I: \,\ SET LEVEL, DISTRIBUTION FUNCTION LEVEL,
QUANTUM LEVEL} \vskip0.5truecm
\begin{tabular}{llll}
\label{TablaI} & \ \ \ \ \ SETS & \ \ FUNCTIONS & \ \ QUANTUM OPERATORS \\
&  &  &  \\
EVOLUTION & $\ \ \ \ A\rightarrow TA$ & $1_{A}\rightarrow
P_{t}1_{A}=1_{T_{A}}$ & $\ \ symb^{-1}1_{T_{A}}=\widehat{P}_{A}(t)$ \\
(projectors) & Liouville ev. & Frob.-Perron ev. & \ \ \ \ Heisenberg ev. \\
&  &  &  \\
EQUILIBRIUM &  & \ \ \ \ $\ U_{t}f_{\ast }=f_{\ast }$ & \ $\ \ \ \ \widehat{U%
}_{t}\widehat{\rho }_{\ast }\widehat{U}_{t}^{\dagger }=\widehat{\rho }_{\ast
}$ \\
(states) &  & \ \ Koopman ev. & \ \ \ Schroedinger ev. \\
&  &  &  \\
OPERATIONS & $\ \ \ \ A\cap B$ & $\ \ \ \ \ \ \ \ \ \ 1_{A}1_{B}$ & \ $\ \ \
\ \ \ \ \ \widehat{P}_{A}\widehat{P}_{B},\hbar \sim 0$ \\
& $\ \ \ \ A\cup B$ & $\ 1_{A}+1_{B}-1_{A}1_{B}$ & $\widehat{P}_{A}+\widehat{%
P}_{B}-\widehat{P}_{A}\widehat{P}_{B}$,$\hbar \sim 0$%
\end{tabular}
\end{itemize}

\section{Ergodic Systems}

According to paper \cite{3M} eq. (E) the system is uniformly ergodic if
\begin{equation}
\lim_{n\rightarrow \infty }\frac{1}{n}\sum_{k=0}^{n-1}\mu (T_{k}B\cap A)=\mu
(A)\mu (B)  \label{E}
\end{equation}%
or%
\begin{equation}
\lim_{n\rightarrow \infty }\frac{1}{n}\sum_{k=0}^{n-1}C(T_{k}B,A)=0
\label{C-E}
\end{equation}%
But if we introduce the measure $\mu _{\ast }(A)$, as we have defined the
new \textit{Ergodic level }making $\mu \rightarrow \mu _{\ast }$, we have that
the system is ergodic if
\begin{equation}
\lim_{n\rightarrow \infty }\frac{1}{n}\sum_{k=0}^{n-1}\mu _{\ast
}(T_{k}A\cap B)=\mu _{\ast }(A)\mu _{\ast }(B)
\end{equation}%
or for the distribution of density function case (see also the corresponding
theorem 4.7 in \cite{Mackeylibro}) or in the continuous case%
\begin{equation*}
\lim_{T\rightarrow \infty }\frac{1}{T}\int_{0}^{T}\langle P_{t}f,g\rangle
dt=\langle f_{\ast },g\rangle
\end{equation*}%
or finally in the quantum case, it is quantum ergodic if
\begin{equation}
\lim_{T\rightarrow \infty }\frac{1}{T}\int_{0}^{T}(\widehat{\rho }(t)|%
\widehat{O})dt=(\widehat{\rho }_{\ast }|\widehat{O})  \label{Q-E}
\end{equation}%
as explained in all details in the first part of this paper i.e. \cite{0}.
We also include the discrete version of the quantum ergodic because in many
situations the evolution of chaotic systems is given in terms of a discrete
evolution operator, for example when the Hamiltonian has a discrete symmetry (see for example the Floquet systems in \cite{stockmann}).

\begin{equation}
\lim_{N\rightarrow \infty }\frac{1}{N}\sum_{k=0}^{N-1}(\widehat{\rho}(k)|
\widehat{O})dt=(\widehat{\rho }_{\ast }|\widehat{O})  \label{Q-E discrete}
\end{equation}

\noindent In section 7 we apply this discrete version to the kicked rotator.

\section{Mixing Systems}

According to paper \cite{3M} eq. (M) the system is \textit{mixing}
if
\begin{equation}
\lim_{n\rightarrow \infty }\mu (T_{n}B\cap A)=\mu (A)\mu (B)  \label{M}
\end{equation}%
or%
\begin{equation}
\lim_{n\rightarrow \infty }C(T_{n}B,A)=0  \label{C-M}
\end{equation}%
Moreover from \cite{Mackeylibro} page 58 the system is \textit{mixing} if
\begin{equation}
\lim_{n\rightarrow \infty }\mu _{\ast }(T_{n}A\cap B)=\mu _{\ast }(A)\mu
_{\ast }(B)
\end{equation}%
or for the distribution of density function case (also see the corresponding
theorem 5.1 in \cite{Mackeylibro})%
\begin{equation*}
\langle P_{t}f,g\rangle =\langle f_{\ast },g\rangle
\end{equation*}%
or in the quantum case, it is quantum mixing if
\begin{equation}
\lim_{t\rightarrow \infty }(\widehat{\rho }(t)|\widehat{O})=(\widehat{\rho }%
_{\ast }|\widehat{O})  \label{quantum mixing condition}
\end{equation}%
as explained in all details in the first part i.e. \cite{0}.

In some cases it might be easier to demonstrate that a system is mixing
using a discrete evolution. The corresponding discrete version of %
\eqref{quantum mixing condition} is

\begin{equation}  \label{quantum mixing condition discrete}
\lim_{N\rightarrow \infty}(\widehat{\rho}(N)|\widehat{O})=(\widehat{\rho}
_{\ast}|\widehat{O})
\end{equation}

\section{Kolmogorov Systems}

The two previous sections are essentially contained in \cite{0} and they
were introduced here for the sake of completeness. This section is the most technical
part of the paper. We remark that things are not so simple at the Kolmogorov
level essentially because the theorem in section 2.4 cannot be reproduced.
We begin by recalling the definition of Kolmogorov systems at the measurable
set level.

\subsection{Kolmogorov systems in the EH}

We return to the definition of Kolmogorov systems of subsection 2.2:

\noindent \textit{A system is Kolmogorov (K-system) if for any integer }$r$%
\textit{\ and any set }$A_{0},A_{1},A_{2},..A_{r} \in X$\textit{\ and for
any }$\varepsilon >0$\textit{\ there exists an }$n_{0}>0$\textit{\ such for
all }$B \in \sigma _{n,r}(A_{1},A_{2},..A_{r})$ \textit{\and} and any $n>n_0$
\textit{\ we have}%
\begin{equation}
|C(B,A_{0})|<\varepsilon
\end{equation}

Then,

\begin{equation}  \label{CondKolmogorov0}
lim_{n\rightarrow \infty }C(B,A_{0})=lim_{n\rightarrow \infty }\{\mu (B\cap
A_{0})-\mu (B)\mu (A_{0})\}=0\text{ \ \ }\forall B\in \sigma
_{n,r}(A_{1},A_{2},...,A_{r})
\end{equation}
where $\sigma _{n,r}(A_{1},A_{2},...,A_{r})$ is the $\sigma $-algebra
generated by $\{T^{k}A_{i}:k\geq n\,\ ;\,\ i=1,...,r\}$, and therefore $%
\sigma _{n,r}(A_{1},A_{2},...,A_{r})=\sigma (\{T^{k}A_{i}:k\geq n\,\ ;\,\
i=1,...,r\})$

Recall that if $f_{\ast }$ is an stationary density, namely $P_{t}f_{\ast
}=f_{\ast }$ then the measure $\mu _{\ast }$ given by

\begin{equation}  \label{medida}
\mu_{*}(A)=\int_{A} f_{*}(\phi)d\phi \,\,\,\,\,\,\,\,\,\,\,\,\ \forall A\in X
\end{equation}

is an invariant measure (i.e. $\mu_{*}(S^{-1}(A))=\mu_{*}(A)$ for all
transformation $S:X\rightarrow X$ and for all $A\in X$) (see Theorem 4.1.1.
of \cite{LM}).

As we consider the previous sections, we make $\mu =\mu _{\ast }$ and
therefore the Kolmogorov condition \eqref{CondKolmogorov0} becomes

\begin{equation}  \label{CondKolmogorov}
lim_{n\rightarrow \infty} \{\mu_{*}(B\cap A_0) - \mu_{*}(B)\mu_{*}(A_0)\} =
0 \,\,\,\,\,\,\,\,\,\,\,\,\ \forall B\in \sigma_{n,r}(A_1,A_2,...,A_r)
\end{equation}

\noindent Now a question arises, What are the sets containing the $\sigma $%
-algebra $\sigma(\{T^{k}A_{i}:k\geq n\,\ ;\,\ i=1,...,r\})?$ There are two
types of these sets:

\begin{enumerate}
\item[$(I)$] $B=\bigcup_{i} T_{n+n_i}A_{s_{i}}\backslash T_{n+l_i}A_{p_{i}}$
\,\,\,\,\,\,\,\,\,\,\,\,\ (finite or countable unions of $T_{i}A_{i}
\backslash T_{j}A_{j}$)

\item[$(II)$] $B=\bigcap_{i} T_{n+n_i}A_{s_{i}}$ \,\,\,\,\,\,\,\,\,\,\,\,\
\,\,\,\,\,\,\,\,\,\,\,\,\,\,\,\,\,\,\,\,\,\,\,\ (finite or countable
intersections of $T_{i}A_{i}$)
\end{enumerate}

\noindent where $n_{i},l_{i}\in \mathbb{N}_{0}$ and $s_{i},p_{i}\in
\{1,...,r\}$.

It is clear that (finite or countable) unions of $T_{i}A_{i}$ are included
because it is sufficient to make in $(I)$ $A_{p_{i}} = \emptyset$ for all $%
p_i$ and results $B=\bigcup T_{n+n_i}A_{s_{i}}$.

Therefore, if we can translate the condition \eqref{CondKolmogorov} into
quantum language for the sets of type $(I)$ and $(II)$ we will have the
Kolmogorov Quantum Hierarchy in the UEH. We begin with the sets of type $(I)$%
:

We have that for these type of sets the condition \eqref{CondKolmogorov}
becomes

\begin{equation}
\begin{split}
&lim_{n\rightarrow \infty} \{\mu_{*}(\bigcup_{i}
T_{n+n_i}A_{s_{i}}\backslash T_{n+l_i}A_{p_{i}}\cap A_0) -
\mu_{*}(\bigcup_{i} T_{n+n_i}A_{s_{i}}\backslash
T_{n+l_i}A_{p_{i}})\mu_{*}(A_0)\} \\
\end{split}%
\end{equation}

\noindent which is equal to

\begin{equation}  \label{Kolmogorov2}
lim_{n\rightarrow \infty} \{\mu_{*}(\bigcup_{i} T_{n+n_i}A_{s_{i}}\cap
(T_{n+l_i}A_{p_{i}})^{c}\cap A_0) - \mu_{*}(\bigcup_{i}
T_{n+n_i}A_{s_{i}}\cap (T_{n+l_i}A_{p_{i}})^{c})\mu_{*}(A_0)\}
\end{equation}

\noindent Now by the \textit{inclusion-exclusion principle} (see for example
\cite{knuth}) if $P$ is a measure of probability and $%
Z_{1},Z_{2},Z_{3},...,Z_{n}$ are sets, then

\begin{equation}
P(\bigcup_{i=1}^{n}Z_{i})=\sum_{k=1}^{n}\sum_{I\subseteq \{1,...,n\},\sharp
(I)=k}(-1)^{k+1}P(\bigcap_{i\in I}Z_{i})
\end{equation}%
\noindent where $P$ is the \textbf{probability} which extended for $%
n\rightarrow \infty $ it becomes

\begin{equation}  \label{exclusion-inclusion}
P(\bigcup_{i=1}^{\infty} Z_i)=\sum_{k=1}^{\infty}\sum_{I\subseteq \mathbb{N}%
, \sharp(I)=k} (-1)^{k+1}P(\bigcap_{i \in I} Z_i)
\end{equation}

\noindent Since that $f_{\ast }$ is a density, more precisely $f_{\ast }\in
D(X,\Sigma ,\mu )=\{f\in L^{1}(X,\Sigma ,\mu ):f\geq 0\,\ ;\,\ \Vert f\Vert
=1\}$ (see Definition 3.1.3. of \cite{LM}), that is, $D(X,\Sigma ,\mu )$ is the
space of the distribution functions defined over all phase space. Then $\mu
_{\ast }$ is a measure of probability and we can use %
\eqref{exclusion-inclusion} to express \eqref{Kolmogorov2} as

\begin{equation}  \label{Kolmogorov3}
\begin{split}
& lim_{n\rightarrow \infty }\sum_{k=1}^{\infty }\sum_{I\subseteq \mathbb{N}%
,\sharp (I)=k}(-1)^{k+1}\mu _{\ast }(\bigcap_{j\in
I}T_{n+n_{j}}A_{s_{j}}\cap (T_{n+l_{j}}A_{p_{j}})^{c}\cap A_{0})- \\
& -lim_{n\rightarrow \infty }\sum_{k=1}^{\infty }\sum_{I\subseteq \mathbb{N}%
,\sharp (I)=k}(-1)^{k+1}\mu _{\ast }(\bigcap_{j\in
I}T_{n+n_{j}}A_{s_{j}}\cap (T_{n+l_{j}}A_{p_{j}})^{c})\mu _{\ast }(A_{0})= \\
& =lim_{n\rightarrow \infty }\sum_{k=1}^{\infty }\sum_{I\subseteq \mathbb{N}%
,\sharp (I)=k}(-1)^{k+1}\{\mu _{\ast }(\bigcap_{j\in
I}T_{n+n_{j}}A_{s_{j}}\cap (T_{n+l_{j}}A_{p_{j}})^{c}\cap A_{0})- \\
& -\mu _{\ast}(\bigcap_{j\in I}T_{n+n_{j}}A_{s_{j}}\cap
(T_{n+l_{j}}A_{p_{j}})^{c})\mu _{\ast }(A_{0})\}= \\
& =lim_{n\rightarrow \infty }\sum_{k=1}^{\infty }\sum_{I\subseteq \mathbb{N}%
,\sharp (I)=k}(-1)^{k+1}C_{\ast}(\bigcap_{j\in I}T_{n+n_{j}}A_{s_{j}}\cap
(T_{n+l_{j}}A_{p_{j}})^{c},A_{0})= \\
& =\sum_{k=1}^{\infty }\sum_{I\subseteq \mathbb{N},\sharp
(I)=k}(-1)^{k+1}lim_{n\rightarrow \infty }C_{\ast}(\bigcap_{j\in
I}T_{n+n_{j}}A_{s_{j}}\cap (T_{n+l_{j}}A_{p_{j}})^{c},A_{0})=0
\end{split}%
\end{equation}

\noindent where we have used $C_{\ast }(A,B)=\mu _{\ast }(A\cap B)-\mu
_{\ast }(A)\mu _{\ast }(B)$. From the last equation \eqref{Kolmogorov3} we
see that the problem reduces to determining if the limit
\begin{equation}
lim_{n\rightarrow \infty }C_{\ast }(\bigcap_{j\in I}T_{n+n_{j}}A_{s_{j}}\cap
(T_{n+l_{j}}A_{p_{j}})^{c},A_{0})=0  \label{Kolmogorov4}
\end{equation}

\noindent exists. So if we translate \eqref{Kolmogorov4} to quantum language
the resultant condition will be the fundamental property of the quantum
Kolmogorov systems (because if we make $A_{p_{i}} = \emptyset$ for all $p_i$
then we obtain the condition of the sets of type $(II)$). In a more general
way we consider infinite numerable intersections

\begin{equation}  \label{Kolmogorov5}
\begin{split}
& lim_{n\rightarrow \infty }C_{\ast}(\bigcap_{j=1}^{\infty
}T_{n+n_{j}}A_{s_{j}}\cap
(T_{n+l_{j}}A_{p_{j}})^{c},A_{0})=lim_{n\rightarrow \infty }\{\mu _{\ast
}(\bigcap_{j=1}^{\infty }T_{n+n_{j}}A_{s_{j}}\cap
(T_{n+l_{j}}A_{p_{j}})^{c}\cap A_{0})- \\
& -\mu _{\ast }(\bigcap_{j=1}^{\infty }T_{n+n_{j}}A_{s_{j}}\cap
(T_{n+l_{j}}A_{p_{j}})^{c})\mu _{\ast }(A_{0})\}=0
\end{split}%
\end{equation}

\noindent Now using the definition of $\mu_*$ (see equation \eqref{medida})
the equation \eqref{Kolmogorov5} is expressed as

\begin{equation}
\begin{split}
&lim_{n\rightarrow \infty }\{\int_{\bigcap_{j=1}^{\infty }
T_{n+n_{j}}A_{s_{j}}\cap (T_{n+l_{j}}A_{p_{j}})^{c}\cap A_{0}}f_{\ast }(\phi
)d\phi - \\
& - (\int_{\bigcap_{j=1}^{\infty } T_{n+n_{j}}A_{s_{j}}\cap
(T_{n+l_{j}}A_{p_{j}})^{c}}f_{\ast }(\phi )d\phi)(\int_{A_o}f_{\ast }(\phi
)d\phi)\}= \\
&=lim_{n\rightarrow \infty }\{\int_{X}1_{\bigcap_{j=1}^{\infty }
T_{n+n_{j}}A_{s_{j}}\cap (T_{n+l_{j}}A_{p_{j}})^{c}\cap A_{0}}f_{\ast }(\phi
)d\phi - \\
& - (\int_{X}1_{\bigcap_{j=1}^{\infty } T_{n+n_{j}}A_{s_{j}}\cap
(T_{n+l_{j}}A_{p_{j}})^{c}}f_{\ast }(\phi )d\phi)(\int_{X}f_{\ast }(\phi
)1_{A_0}d\phi)\}=0 \\
\end{split}%
\end{equation}

\noindent which is equal to

\begin{equation}  \label{Kolmogorov6}
\begin{split}
&lim_{n\rightarrow \infty }\{
\int_{X}\prod_{j=1}^{%
\infty}1_{T_{n+n_{j}}A_{s_{j}}}(1-1_{T_{n+l_{j}}A_{p_{j}}})f_{\ast }(\phi
)1_{A_0}d\phi- \\
&-(\int_{X}\prod_{j=1}^{%
\infty}1_{T_{n+n_{j}}A_{s_{j}}}(1-1_{T_{n+l_{j}}A_{p_{j}}})f_{\ast }(\phi
)d\phi)(\int_{X}f_{\ast }(\phi )1_{A_0}d\phi)\}=0
\end{split}%
\end{equation}

\noindent Moreover the characteristic functions $%
1_{T_{n+n_{j}}A_{s_{j}}},1_{T_{n+l_{j}}A_{p_{j}}}$ are equal to $%
P_{n+n_{j}}1_{A_{s_{j}}},P_{n+l_{j}}1_{A_{p_{j}}}$ respectively. Then %
\eqref{Kolmogorov6} becomes

\begin{equation}  \label{Kolmogorov7}
\begin{split}
&lim_{n\rightarrow \infty }\{
\int_{X}\prod_{j=1}^{%
\infty}P_{n+n_{j}}1_{A_{s_{j}}}(1-P_{n+l_{j}}1_{A_{p_{j}}})f_{\ast }(\phi
)1_{A_0}d\phi- \\
&-(\int_{X}\prod_{j=1}^{%
\infty}P_{n+n_{j}}1_{A_{s_{j}}}(1-P_{n+l_{j}}1_{A_{p_{j}}})f_{\ast }(\phi
)d\phi)(\int_{X}f_{\ast }(\phi )1_{A_0}d\phi)\}=0
\end{split}%
\end{equation}

\noindent Using that

\begin{equation}
1-P_{n+l_{j}}1_{A_{p_{j}}}=P_{n+l_{j}}(1-1_{A_{p_{j}}})=P_{n+l_{j}}1_{(A_{p_{j}})^{c}}
\end{equation}

\noindent we see that the equation \eqref{Kolmogorov7} can be expressed as

\begin{equation}
\begin{split}
&lim_{n\rightarrow \infty }\{
\int_{X}\prod_{j=1}^{%
\infty}P_{n+n_{j}}1_{A_{s_{j}}}P_{n+l_{j}}1_{(A_{p_{j}})^{c}}f_{\ast}(\phi
)1_{A_0}d\phi- \\
&-(\int_{X}\prod_{j=1}^{%
\infty}P_{n+n_{j}}1_{A_{s_{j}}}P_{n+l_{j}}1_{(A_{p_{j}})^{c}}f_{\ast }(\phi
)d\phi)(\int_{X}f_{\ast }(\phi )1_{A_0}d\phi)\}=0
\end{split}%
\end{equation}

\noindent and therefore with the same trick used in subsection 2.4 we have

\begin{equation}
\begin{split}
&lim_{n\rightarrow \infty }\{ \langle
\prod_{j=1}^{%
\infty}P_{n+n_{j}}1_{A_{s_{j}}}P_{n+l_{j}}1_{(A_{p_{j}})^{c}}f_{\ast
}(\phi),1_{A_0}\rangle- \\
&-\langle
\prod_{j=1}^{%
\infty}P_{n+n_{j}}1_{A_{s_{j}}}P_{n+l_{j}}1_{(A_{p_{j}})^{c}}f_{\ast }(\phi
),1\rangle\langle f_{\ast }(\phi ),1_{A_0}\rangle\}=0
\end{split}%
\end{equation}

\noindent Then, we consider three sets of generic numbers $(a_{n}^{(j)})$, $%
(b_{m}^{(j)})$ and $(c_{l})$ and we define the generic functions

\begin{equation}
\begin{split}
&f_{s_j}=\sum_{n}a_n^{(j)} 1_{A_{s_{j}}^{(n)}} \\
&f_{p_j}=\sum_{m}b_m^{(j)} 1_{((A_{p_{j}})^{c})^{(m)}}f_{*}(\phi) \\
&g=\sum_{l}c_l 1_{A_{0}^{(l)}}
\end{split}%
\end{equation}

\noindent Then we obtain that

\begin{equation}
\begin{split}
&lim_{n\rightarrow \infty }\{ \langle
\prod_{j=1}^{\infty}P_{n+n_{j}}f_{s_{j}}P_{n+l_{j}}f_{p_{j}},g\rangle- \\
&-\langle
\prod_{j=1}^{\infty}P_{n+n_{j}}f_{s_{j}}P_{n+l_{j}}f_{p_{j}},1\rangle\langle
f_{\ast }(\phi ),g\rangle\}=0
\end{split}%
\end{equation}

\noindent If we realign the indices $n_{j},l_{j}$ and define the functions $%
f_{s_{j}},f_{p_{j}}$ such that $n_{j}=m_j,l_{j}=m_{j+1}$; $%
F_{j}=f_{s_{j}},F_{j+1}=f_{p_{j}}$, that is, $f_{s_{j}}$ and $f_{p_{j}}$ are
the $F_j$ terms of even and odd index. We have

\begin{equation}  \label{Kolmogorov8}
lim_{n\rightarrow \infty }\{ \langle
\prod_{j=1}^{\infty}P_{n+m_{j}}F_{j},g\rangle -\langle
\prod_{j=1}^{\infty}P_{n+m_{j}}F_{j},1\rangle\langle f_{\ast}(\phi
),g\rangle\}=0
\end{equation}

\noindent We can rewrite \eqref{Kolmogorov8} as

\begin{equation}  \label{Kolmogorov9}
lim_{n\rightarrow \infty}\{\langle
P_{n+m_{1}}F_{1},g\prod_{j=2}^{\infty}P_{n+m_{j}}F_{j}\rangle\ -\langle
P_{n+m_{1}}F_{1},\prod_{j=2}^{\infty}P_{n+m_{j}}F_{j}\rangle\ \langle
f_{\ast}(\phi ),g\rangle\}=0
\end{equation}

\noindent Now according to paper \cite{0} the star product tends to the product
function when $\hbar \rightarrow 0$

\begin{equation}  \label{Kolmogorov10}
f(\phi)g(\phi) \longrightarrow (f \ast g)(\phi) = symb(\widehat{f}) \ast
symb(\widehat{g}) = symb(\widehat{f}\widehat{g})
\end{equation}

\noindent and therefore when $\hbar \rightarrow 0$ for an infinite product
of functions we have

\begin{equation}  \label{symb product}
\begin{split}
& f_i(\phi) = symb(\widehat{f_i}) \\
& \prod_{i=1}^{\infty} f_i(\phi) = \prod_{i=1}^{\infty} symb(\widehat{f_i})
= symb(\prod_{i=1}^{\infty} \widehat{f_i})
\end{split}%
\end{equation}

\noindent On the other hand from table 1 (subsection 2.4) we have

\begin{equation}
P_t 1_{A} = 1_{T_t A}=symb(\widehat{P}_{A}(t))
\end{equation}

\noindent and therefore, if we have a generic function $h$

\begin{equation}
h = \sum_{k} h_k 1_{A_k}
\end{equation}

\noindent then (see eq. (22) of paper \cite{0} and table 1 of subsection 2.4)

\begin{equation}
\begin{split}
& symb^{-1}(1_{A_k}) = \widehat{P}_{A_k} \\
& \widehat{h} = symb^{-1}(h) = \sum_{k} h_k symb^{-1}(1_{A_k})=\sum_{k} h_k
\widehat{P}_{A_k} \\
& \widehat{h}(t) = \sum_{k} h_k \widehat{P}_{A_k}(t) \\
& P_t h = \sum_{k} h_k P_t 1_{A_k} = \sum_{k} h_k P_t 1_{A_k} = \sum_{k} h_k
P_t 1_{A_k} = \sum_{k} h_k symb(\widehat{P}_{A_k}(t)) = \\
& = symb(\sum_{k} h_k \widehat{P}_{A_k}(t)) = symb(\widehat{h}(t))
\end{split}%
\end{equation}

\noindent Then if we introduce the last equation of (58) and the equation
(55) $P_{n+m_{j}}F_{j}$ (when $\hbar \rightarrow 0)$ becomes

\begin{equation}  \label{symb product1}
\begin{split}
& P_{n+m_j}F_j = symb(\widehat{F}_j(n+m_j)) \\
& \prod_{j=2}^{\infty}P_{n+m_{j}}F_{j} = \prod_{j=2}^{\infty}symb(\widehat{F}%
_j(n+m_j)) = symb(\prod_{j=2}^{\infty}\widehat{F}_j(n+m_j))
\end{split}%
\end{equation}

\noindent Now we call

\begin{equation}  \label{symb product2}
\begin{split}
& f_{\ast} = symb(\widehat{\rho_{\ast}}) \\
& g = symb(\widehat{g})
\end{split}%
\end{equation}

\noindent where $\widehat{\rho}_{\ast}$ is the weak limit of $\widehat{\rho}%
(t)$ (see \cite{0}).

Therefore if we use \eqref{symb product}, \eqref{symb product1} and %
\eqref{symb product2} in \eqref{Kolmogorov9} when $\hbar \rightarrow 0$ we
have

\begin{equation}
\begin{split}
&lim_{n\rightarrow \infty }\{ \langle symb(\widehat{F}_{1}(n+m_{1})),symb(%
\widehat{g}\prod_{j=2}^{\infty}\widehat{F}_{j}(n+m_{j}))\rangle \\
& -\langle symb(\widehat{F}_{1}(n+m_{1})),symb(\prod_{j=2}^{\infty}\widehat{F%
}_{j}(n+m_{j}))\rangle\langle symb(\rho_{\ast}),symb(\widehat{g})\rangle\}=0
\end{split}%
\end{equation}

\noindent Now this equation can be expressed in the quantum level as
(replacing $\langle\,\ , \,\ \rangle$ by $(\,\ | \,\, )$)

\begin{equation}
\begin{split}
&lim_{n\rightarrow \infty }\{ ( symb(\widehat{F}_{1}(n+m_{1}))|symb(\widehat{%
g}\prod_{j=2}^{\infty}\widehat{F}_{j}(n+m_{j}))) \\
& -( symb(\widehat{F}_{1}(n+m_{1}))|symb(\prod_{j=2}^{\infty}\widehat{F}%
_{j}(n+m_{j}))) ( symb(\rho_{\ast})|symb(\widehat{g}))\}=0
\end{split}%
\end{equation}

\noindent At this point we rename the operators $\widehat{F}_{1}(n+m_{1}),$
as $\widehat{F}_{j}(n+m_{j})$ and $\widehat{g}$ as $\widehat{\rho }(n+m_{1}),%
\widehat{O}_{j}(n+m_{j})$ and $\widehat{O}_{1}$ respectively. That is, if we
emphasize the role of the states and the observables we have

\begin{equation}
\begin{split}
&lim_{n\rightarrow \infty }\{ ( symb(\widehat{\rho}(n+m_{1}))|symb(\widehat{O%
}_1\prod_{j=2}^{\infty}\widehat{O}_{j}(n+m_{j}))) \\
& -( symb(\widehat{\rho}(n+m_{1}))|symb(\prod_{j=2}^{\infty}\widehat{O}%
_{j}(n+m_{j}))) ( symb(\rho_{\ast})|symb(\widehat{O}_1))\}=0
\end{split}%
\end{equation}

\noindent Then, using the important property that the Wigner transformation
yields the correct expectation value of any observable $\widehat{O}$ in the
state $\widehat{\rho }$ (see equation (23) of paper \cite{0}) we have

\begin{equation}
lim_{n\rightarrow \infty}\{(\widehat{\rho}(n+m_1)|\widehat{O}%
_1\prod_{j=2}^{\infty}\widehat{O}_j(n+m_j))-\prod_{j=2}^{\infty}(\widehat{%
\rho}(n+m_1)|\widehat{O}_j(n+m_j))(\widehat{\rho_{\ast}}|\widehat{O}_1)\}=0
\end{equation}

\noindent Finally, the definition of the quantum Kolmogorov level is

\begin{equation}  \label{quantum Kolmogorov condition}
lim_{n\rightarrow \infty}\{(\widehat{\rho}(n+m_1)|\widehat{O}%
_1\prod_{j=2}^{\infty}\widehat{O}_j(n+m_j))-\prod_{j=2}^{\infty}(\widehat{%
\rho}(n+m_1)|\widehat{O}_j(n+m_j))(\widehat{\rho_{\ast}}|\widehat{O}_1)\}=0
\end{equation}

\noindent for all observables $\widehat{O}_{2},\widehat{O}_{3},\widehat{O}%
_{4},...$ and all $m_1, m_2,m_3,...\in \mathbb{N}_{0}$ where $\widehat{%
\rho_{\ast}}$ is the weak limit of $\widehat{\rho}(t)$.

\subsection{Particular Case: Mixing}

According to the definition of Kolmogorov level, we know that the mixing
level includes Kolmogorov level, that is, the equation %
\eqref{CondKolmogorov0} implies the equation \eqref{M}. Therefore it would
be expected for a good definition of quantum Kolmogorov level given by
equation \eqref{quantum Kolmogorov condition} that from this equation we can
deduce the quantum level mixing given by equation
\eqref{quantum mixing
condition}. If we make $O_1=\widehat{O}$, $O_i=\widehat{I}$ for all
i=2,3,4... and $m_1=0$ in the equation (65) we have

\begin{equation}
lim_{n\rightarrow \infty}\{(\widehat{\rho}(n)|\widehat{O})-(\widehat{\rho}%
(n)|\widehat{I})(\widehat{\rho_{\ast}}|\widehat{O})\}=0
\end{equation}

\noindent and since $(\widehat{\rho}(n)|\widehat{I})=Tr(\widehat{\rho}%
(n))=Tr(\widehat{\rho}(0))=1$ (conservation of the trace given by equation %
\eqref{C}) we have

\begin{equation}
lim_{n\rightarrow \infty}\{(\widehat{\rho}(n)|\widehat{O})-(\widehat{%
\rho_{\ast}}|\widehat{O})\}=0
\end{equation}

\noindent Then,

\begin{equation}
lim_{n\rightarrow \infty}(\widehat{\rho}(n)|\widehat{O})=(\widehat{%
\rho_{\ast}}|\widehat{O})
\end{equation}

\noindent which is identical to the limit

\begin{equation}
lim_{t\rightarrow \infty}(\widehat{\rho}(t)|\widehat{O})=(\widehat{%
\rho_{\ast}}|\widehat{O})
\end{equation}

\noindent That is, $\widehat{\rho}(t)$ weakly converges to $\widehat{%
\rho_{\ast}}$ corresponding to the mixing case. Therefore, the quantum
Kolmogorov level implies the quantum mixing level.

\section{Bernoulli Systems}

Essentially the Bernoulli systems satisfy the mixing conditions but with no $%
\lim_{t\rightarrow \infty }$ $\ $(see e.g. eqs. (\ref{MI}) and (\ref{BE})).
Then these systems satisfy the following equations:

According to paper \cite{3M} eq. (BE) the system is \textit{Bernoulli} if
\begin{equation}\label{Bernoulli}
\mu (T_{n}B\cap A)=\mu (A)\mu (B)
\end{equation}
\noindent or
\begin{equation}
C(T_{n}B,A)=0
\end{equation}
\noindent or for the distribution of density function case (see also the
corresponding theorem in \cite{M})
\begin{equation*}
\langle P_{t}f,g\rangle =\langle f_{\ast },g\rangle
\end{equation*}
\noindent or in the quantum case, it is quantum Bernoulli if
\begin{equation}  \label{Bernoulli condition}
(\widehat{\rho }(t)|\widehat{O})=(\widehat{\rho }_{\ast }|\widehat{O}).
\end{equation}

\noindent Since the Bernoulli condition \eqref{Bernoulli condition} is
independent of time then it becomes unnecessary to have a discrete version
of this condition.

\subsection{Independent Events}

Let $A\subseteq X$ be an event of the phase space. If we interpret $\mu (A)$
as the probability $P(A)$ of $A$, then Bernoulli systems satisfy
a property expressing the independence between two events of the phase
space. This property follows directly from its definition. Let $A$ and $B$
be two events, then if $n=0$ in the equation %
\eqref{Bernoulli} we have the \emph{independence events property}:

\begin{equation}  \label{independence}
\mu (B\cap A)=\mu (A)\mu (B)
\end{equation}

\noindent I.e., the probability that $A$ and $B$ occurs simultaneously is the
product of the probability of $A$ by the probability of $B$.

Now if we take $\mu_{\ast}=\mu$ with $\mu_{\ast}(A)=\int_{A}f_{\ast}(\phi)d%
\phi$ then

\begin{equation}
\begin{split}
& \mu_{\ast} (B\cap A)=\mu_{\ast} (A)\mu_{\ast} (B) \\
& \int_{A\cap
B}f_{\ast}(\phi)d\phi=\int_{A}f_{\ast}(\phi)d\phi\int_{B}f_{\ast}(\phi)d\phi
\end{split}%
\end{equation}

\noindent Namely,

\begin{equation}  \label{bernoulli1}
\begin{split}
& \int_{X}f_{\ast}1_A 1_B d\phi=\int_{X}f_{\ast}1_A d\phi\int_{X}f_{\ast}1_B
d\phi \\
& \langle f_{\ast},1_A 1_B\rangle=\langle f_{\ast},1_A\rangle \langle
f_{\ast},1_B\rangle
\end{split}%
\end{equation}

\noindent Let $g_{1}=\sum_{k}a_{k}1_{A_{k}}$ and $g_{2}=%
\sum_{l}b_{l}1_{B_{l}}$ be. From \eqref{bernoulli1} we have

\begin{equation}
a_k b_l\langle f_{\ast},1_{A_k} 1_{B_l}\rangle=a_k b_l\langle
f_{\ast},1_{A_k}\rangle\langle f_{\ast},1_{B_l}\rangle
\end{equation}

\noindent By the linearity of the inner product and summing over the indices
k and l we have

\begin{equation}
\langle f_{\ast},\sum_{k}a_k 1_{A_k} \sum_{l}b_l 1_{B_l}\rangle=\langle
f_{\ast},\sum_{k} a_k 1_{A_k}\rangle \langle f_{\ast},\sum_{l}b_l
1_{B_l}\rangle
\end{equation}

\noindent That is,

\begin{equation}
\langle f_{\ast},g_1 g_2\rangle=\langle f_{\ast},g_1\rangle \langle
f_{\ast},g_2\rangle
\end{equation}

\noindent Therefore, if $f_{\ast}=symb(\widehat{\rho_{\ast}})$ and $g_1=symb(%
\widehat{g_1}),g_2=symb(\widehat{g_2})$ where $\widehat{g_1}$ and $\widehat{%
g_2}$ are observables we have

\begin{equation}
\langle symb(\widehat{\rho_{\ast}}),g_1 g_2\rangle=\langle symb(\widehat{%
\rho_{\ast}}),symb(\widehat{g_1})\rangle \langle symb(\widehat{\rho_{\ast}}%
),symb(\widehat{g_2})\rangle
\end{equation}

\noindent Now from $g_{1}(\phi )g_{2}(\phi )\rightarrow symb(\widehat{g_{1}}%
\widehat{g_{2}})$ when $\hbar \rightarrow 0$ (see equation %
\eqref{Kolmogorov10}) we obtain

\begin{equation}
\langle symb(\widehat{\rho_{\ast}}),symb(\widehat{g_1}\widehat{g_2}%
)\rangle=\langle symb(\widehat{\rho_{\ast}}),symb(\widehat{g_1})\rangle
\langle symb(\widehat{\rho_{\ast}}),symb(\widehat{g_2})\rangle
\end{equation}

\noindent Namely,

\begin{equation}  \label{bernoulli2}
(\widehat{\rho_{\ast}}|\widehat{g_1}\widehat{g_2})=(\widehat{\rho_{\ast}}|%
\widehat{g_1})(\widehat{\rho_{\ast}}|\widehat{g_2})
\end{equation}

\noindent where in \eqref{bernoulli2} we have used the fundamental property
of the \emph{symb} given by the equation (24) of \cite{0}. Moreover, we know
that

\begin{equation}
\begin{split}
& (\widehat{\rho}(t)|\widehat{g_1}\widehat{g_2})=(\widehat{\rho_{\ast}}|%
\widehat{g_1}\widehat{g_2}) \\
& (\widehat{\rho}(t)|\widehat{g_1})=(\widehat{\rho_{\ast}}|\widehat{g_1}) \\
& (\widehat{\rho}(t)|\widehat{g_2})=(\widehat{\rho_{\ast}}|\widehat{g_2})
\end{split}%
\end{equation}

\noindent Therefore we can express \eqref{bernoulli2} as

\begin{equation}  \label{bernoulli3}
(\widehat{\rho}(t)|\widehat{g_1}\widehat{g_2})=(\widehat{\rho}(t)|\widehat{%
g_1})(\widehat{\rho}(t)|\widehat{g_2})
\end{equation}

\noindent for all pairwise of observables$,\widehat{g_{1}}\widehat{g_{2}}$.
If we generalize for an arbitrary product of observables from %
\eqref{bernoulli3} we have that

\begin{equation}  \label{bernoulli4}
(\widehat{\rho}(t)|\prod_{i}\widehat{g_i})=\prod_{i}(\widehat{\rho}(t)|%
\widehat{g_i})
\end{equation}

\noindent The equation \eqref{bernoulli4} is the translation into quantum
language of the independence of events expressed by the equation %
\eqref{independence}. Physically, it tells us that, in the classical limit
of a Bernoulli system the mean value of an arbitrary product of observables
factorizes into the product of the mean values of each observable and this
factorization occurs at all times, and this quantum factorization express
the no-correlation of the observables $\widehat{g_{1}},\widehat{g_{2}},..$
of (87) in a Bernouilli system.

\subsection{Particular Case: Kolmogorov}

Bernoulli level is included in Kolmogorov level (equation \eqref{Bernoulli}). This fact implies equation \eqref{CondKolmogorov} so this property must
be verified by the respective quantum versions on these levels. We consider
a numerable set of observables $\widehat{O}_{1},\widehat{O}_{2},\widehat{O}%
_{3},...$ and a sequence $m_{1},m_{2},m_{3}...\in \mathbb{N}_{0}$. To
demonstrate that quantum Bernoulli level implies quantum Kolmogorov level we
use the quantum version of the \emph{independence events property} given by
the equation \eqref{bernoulli4}. If we call $\widehat{g}_{1}=\widehat{O}_{1}$%
, $\widehat{g}_{2}=\prod_{j=2}^{\infty }\widehat{O}_{j}(n+m_{j})$ for all $%
j=2,3,4,...$ by the equation \eqref{bernoulli4} we have

\begin{equation}  \label{bernoulli5}
(\widehat{\rho}(n+m_1)|\widehat{O}_1\prod_{j=2}^{\infty}\widehat{O}%
_j(n+m_j))=(\widehat{\rho}(n+m_1)|\widehat{O}_1)\prod_{j=2}^{\infty}(%
\widehat{\rho}(n+m_1)|\widehat{O}_j(n+m_j))
\end{equation}

\noindent In particular since the system is Bernoulli we have

\begin{equation}  \label{bernoulli6}
(\widehat{\rho}(n+m_1)|\widehat{O}_1)=(\widehat{\rho_{\ast}}|\widehat{O}_1)
\end{equation}

\noindent From equations \eqref{bernoulli5} and \eqref{bernoulli6} we have
that

\begin{equation}
(\widehat{\rho}(n+m_1)|\widehat{O}_1\prod_{j=2}^{\infty}\widehat{O}%
_j(n+m_j))=\prod_{j=2}^{\infty}(\widehat{\rho}(n+m_1)|\widehat{O}_j(n+m_j))(%
\widehat{\rho_{\ast}}|\widehat{O}_1)
\end{equation}

\noindent Therefore,

\begin{equation}
lim_{n\rightarrow \infty}(\widehat{\rho}(n+m_1)|\widehat{O}%
_1\prod_{j=2}^{\infty}\widehat{O}_j(n+m_j))=lim_{n\rightarrow
\infty}\prod_{j=2}^{\infty}(\widehat{\rho}(n+m_1)|\widehat{O}_j(n+m_j))(%
\widehat{\rho_{\ast}}|\widehat{O}_1)
\end{equation}

\noindent That is,

\begin{equation}
lim_{n\rightarrow \infty}\{(\widehat{\rho}(n+m_1)|\widehat{O}%
_1\prod_{j=2}^{\infty}\widehat{O}_j(n+m_j))-\prod_{j=2}^{\infty}(\widehat{%
\rho}(n+m_1)|\widehat{O}_j(n+m_j))(\widehat{\rho_{\ast}}|\widehat{O}_1)\}=0
\end{equation}

\noindent which is the quantum Kolmogorov condition (see equation %
\eqref{quantum Kolmogorov condition}).

\section{Physical Relevance of QEH: Casati-Prosen model and Kicked Rotator}

In this section we give a physical relevance to the Quantum Ergodic
Hierarchy analyzing in terms of the QEH levels two standard models of the
quantum chaos literature: the Casati-Prosen model \cite{casati verdadero} and
the kicked rotator (\cite{stockmann}, \cite{casati libro}, \cite{haake}). These models are emblematic for quantum chaos because
the first one contains the mean features of the chaotic billiards which
are one of the pioneers experiments made in the field (\cite{stockmann}, \cite{casati libro}) and the two others contains
the physics of many relevant phenomenons (see \cite{stockmann} chapter 4.2) like the Anderson localization, the hydrogen atom in an electric field, etc.

\subsection{The Casati-Prosen model in the quasi-continuous spectrum
approximation}

Casati-Prosen model is a Sinai billiard provided with a quantum paradigmatic
phenomenon - the double slit experiment. On the other hand the classical
Sinai billiards are a wellknown type of Kolmogorov systems and therefore
they are mixing systems. In this subsection we review the quasi-continuous
spectrum approximation we used in \cite{casati model} to explain the
Casati-Prosen model conceptually. We briefly begin summarizing these
arguments:

\begin{itemize}
\item For large times but shorter than the time \footnote{Strictly, $t^*$ is an approximation to the Poincare time \cite{POINCARE} since $e^{-i(\frac{E_n-E_m}{\hbar})t^*}=e^{-2\pi i\frac{E_n-E_m}{min{|E_{i}-E_{i+1}|}}}\approx 1$ if and only if $\frac{E_n-E_m}{min{|E_{i}-E_{i+1}|}}$ is an integer for all $n,m$ but in general this requirement can be fulfilled for some initial conditions $\rho(0)$. To avoid these problems we indeed should use $max{|E_{i}-E_{i+1}|}$ instead of $min{|E_{i}-E_{i+1}|}$ for the quasi-continuous approximation. If we put $t^*=\frac{2\pi\hbar}{max{|E_{i}-E_{i+1}|}}$ since $max\{|E_{i}-E_{i+1}|\}\geq \Delta E$ where $\Delta E$ is mean energy level spacing it follows that $t^*\leq t_H$ where $t_H=\frac{2\pi\hbar}{\Delta E}$ is the Heisenberg time \cite{HEISENBERG}. This hypothesis is reasonable because quantum chaos phenomena with a semiclassical description such as relaxation, exponential sensitivity etc. are possible within a time scale $t\leq t^*$.
    Then, the quasi continuous spectrum approximation is obtained as follows: if $t\ll t^*$ for all finite times $t$ $\Longrightarrow  t^*=\frac{2\pi\hbar}{max{|E_{i}-E_{i+1}|}}\rightarrow \infty \Longrightarrow \frac{max|E_i-E_{i+1}|}{2\pi\hbar}\approx0 \Longrightarrow \frac{|E_i-E_{i+1}|}{2\pi\hbar}\approx0$ then $\frac{E_i-E_{i+1}}{\hbar}$ is infinitesimal. Therefore, we can replace any sum $\sum_j f(E_j)e^{-i\frac{E_j}{\hbar}t}$ by the integral $\int dEf(E)e^{-i\frac{E}{\hbar}t}$. Then when $t\ll t^*$ the Riemann-Lebesgue theorem can be used.} $t^*=\frac{2\pi\hbar}{min{|E_{i}-E_{i+1}|}}$ of the
quantum system, that is $t\ll t^*$, we can assume that the energy spectrum is quasi continuous and therefore
we can replace sums by integrals in all the equations. At this point we
consider the system is mixing (moreover it is a K-system) and then due to the
decoherence of the Sinai billiard the interference is calculated in the
equilibrium state $\widehat{\rho }_{\ast }$ which is the weak limit of the
initial state $\widehat{\rho }(0)$ (mixing condition).

\item To define the nonintegrability of the Sinai billiard we use the local
CSCO of \cite{0} $\{\widehat{H},\widehat{O}_{\phi _{i}}\}$. For this case $%
\widehat{O}_{\phi _{i}}=\widehat{P}_{\phi _{i}}$ is the local momentum and $%
\widehat{P}_{\phi _{i}}=(\widehat{P}_{x},\widehat{P}_{y})_{\phi _{i}}$.

\item In these terms, the interference $P_{int}$ of the state $\rho
(t)=|\varphi (x,t)\rangle \langle \varphi (x,t)|$ (where $|\varphi
(x,t)\rangle =|\varphi _{1}(x,t)\rangle +|\varphi _{2}(x,t)\rangle $ and $%
|\varphi _{i}(x,t)\rangle $ are the two circular-symmetric solutions
produced by the boundary conditions in the two slits as a result of the
direct impact of the initial gaussian wavepacket) it is given by $2Re(\varphi
_{1}(x,t)\varphi _{2}(x,t)^{\ast })$.

\item The state $\rho (t)=|\varphi (x,t)\rangle \langle \varphi (x,t)|$ is
replaced by the equilibrium state $\widehat{\rho }_{\ast }$ where the time
dependence has disappeared in the equilibrium. As a consequence of the
linearity of the Schrodinger equation and by the local CSCO $\{\widehat{H},%
\widehat{P}_{\phi _{i}}\}$ the interference is $P_{int}=2\sum_{ipw}\rho
(w)_{\phi _{i},p}Re(\varphi _{1wp}(x)\varphi _{2wp}(x)^{\ast })$.

\item To compute $P_{int}$ we use the unitary transform $U_{p\phi}^{m}$ that
diagonalizes $(w,m,m^{\prime}|_{\phi}$ so $P_{int}$ is

\begin{equation}\label{7.1}
P_{int}=\sum_{i,p,p^{\prime},w,m,m^{\prime}}\rho(w)_{\phi_i,p,p^{%
\prime}}[U_{p\phi_i}^{m}(U_{p^{\prime}\phi_i}^{m^{\prime}})^{\ast} e^{-\frac{
i}{\hbar}(\mathbf{m}-\mathbf{m^{\prime}}).\mathbf{x}}e^{-\frac{i}{\hbar}(
\frac{\mathbf{m}+\mathbf{m^{\prime}}}{2}).\mathbf{s}} +C.C.]
\end{equation}
\noindent where C.C. denotes the complex conjugate and $\mathbf{s}=(s,0)$
where $s$ is the distance between the slits. The set $\{w\}$ is the energy
spectrum which we assume quasi-continuous and $\mathbf{m}=\hbar \mathbf{k}$ labels each wave vector in the expansion.

\item Now because there is a macroscopic distance from the two slits screen
to the photographic plate, $\mathbf{x}$ is macroscopic with respect to $
\hbar $ in such a way that we can consider that $\frac{x}{\hbar}
\rightarrow\infty$ and we can use the Riemann-Lebesgue theorem concluding
that $P_{int}=0$ \footnote{Note: $\mathbf{x}$ and $\hbar$ have not the same units and therefore the limit $\frac{x}{\hbar}
\rightarrow\infty$ has the following explanation. Since $\mathbf{m}=\hbar\mathbf{k}$ the factor $e^{-\frac{
i}{\hbar}(\mathbf{m}-\mathbf{m^{\prime}}).\mathbf{x}}$ in \eqref{7.1} is equal to $e^{-i(\mathbf{k}-\mathbf{k^{\prime}}).\mathbf{x}}$. Further, because $E=\frac{\hbar \mathbf{k}^2}{2M}$ in the quasi-continuous spectrum approximation, $\mathbf{k}$ is quasi-continuous. Moreover, since
$|\mathbf{k}|=\frac{2\pi}{\lambda}$ and $\mathbf{x}$ is macroscopic with respect to $\lambda$ (for example, in a electron wavepacket $\lambda$ is typically of the order of $10^{-13}$ cm) we can consider that $\frac{2\pi\mathbf{x}}{\lambda}=\mathbf{k}.\mathbf{x}\rightarrow \infty$.
Now we can we can use the Riemann-Lebesgue theorem in the sum of eq. \eqref{7.1} concluding that $P_{int}=0$.}.
\end{itemize}

\noindent In this way in \cite{casati model} we demonstrate that the
interference fringes vanish due to the decoherence in the equilibrium state.
From the viewpoint of the Quantum Ergodic Hierarchy the equilibrium state $
\widehat{\rho }_{\ast }$ of the Casati model is a consequence that this
model belongs to the mixing level. And in turn under the hypothesis
mentioned above the mixing of Casati-Prosen model implies the cancellation
of the interference fringes. On the other hand, Casati explains his model
\cite{casati verdadero} by a computer experiment that shows how complexity
can produce this decoherence. Therefore the relevant observation is that the
Quantum Ergodic Hierarchy brings us a conceptual framework where
computability can be tested and moreover, this test is in agreement with the
theoretical results of QEH \cite{casati model}. This is a proof of the
physical relevance to QEH.

\subsection{The Explanation Of Casati-Prosen model in terms of the ergodic
level}

From all above arguments mentioned in the section 7.1 two \textquotedblleft
natural" objections can be made: Since most of chaotic systems of interest
are bounded systems with energy spectrum discrete, can we be sure that
hypothesis quasi-continuous spectrum is valid for all these systems? And
also, since in all expressions of the QEH levels the time goes to infinity,
What happens when the time $t^{*}=\frac{2\pi\hbar }{min{
|E_{i}-E_{i+1}|}}$ is small?

The answer to the first question is negative, and in such a case we can
still explain these systems in terms of QEH. In the next subsection and
below we show this fact and also this is the answer to the second objection.
Therefore, we end this subsection giving an alternative explanation. That
is, we strictly assume that the energy spectrum of the Casati-Prosen model
is discrete. This is so because the Casati-Prosen model is a bounded system
and therefore the energy spectrum is discrete.

First, due to that the Casati-Prosen model is a Kolmogorov then it is
ergodic. In particular, the equilibrium state $\widehat{\rho }_{\ast }$ is
the Cèsaro limit of the initial Gaussian wavepacket $\widehat{\rho }(0)$. On
the other hand the initial state written in terms of the local CSCO $\{%
\widehat{H},\widehat{P}_{\phi _{i}}\}$ is given by \cite{0}

\begin{equation}
\widehat{\rho}(0)=\sum_{i,p,p^{\prime},w_{\alpha},w_{\alpha}^{\prime}}%
\rho_{w_{\alpha},w_{\alpha}^{\prime},\phi_i,p,p^{\prime}}
U_{p\phi_i}^{m}(U_{p^{\prime}\phi_i}^{m^{\prime}})^{\ast}|w_{\alpha},m%
\rangle_{\phi_i}\langle w_{\alpha}^{\prime},m^{\prime}|_{\phi_i}
\end{equation}

\noindent Then the state at the time $t$ is

\begin{equation}
\widehat{\rho}(t)=\sum_{i,p,p^{\prime},w_{\alpha},w_{\alpha}^{\prime}}%
\rho_{w_{\alpha},w_{\alpha}^{\prime},\phi_i,p,p^{\prime}}U_{p%
\phi_i}^{m}(U_{p^{\prime}\phi_i}^{m^{\prime}})^{\ast}e^{-\frac{i}{\hbar}%
(w_{\alpha}-w_{\alpha}^{\prime})t}|w_{\alpha},m\rangle_{\phi_i}\langle
w_{\alpha}^{\prime},m^{\prime}|_{\phi_i}
\end{equation}

\noindent We are interested in the probability amplitude $|\varphi(\mathbf{x}%
,t)|^{2}$ (see eq. (6) of \cite{casati model}) at $\mathbf{x}$ for the state
$\widehat{\rho}(t)$ in the limit $t\rightarrow \infty$. Then this amplitude
is given by
\begin{equation}  \label{amplitud cesaro prosen1}
\langle \mathbf{x}|\widehat{\rho}(t)|\mathbf{x^{\prime}}\rangle=%
\sum_{i,p,p^{\prime},w_{\alpha},w_{\alpha}^{\prime}}\rho_{w_{\alpha},w_{%
\alpha}^{\prime},\phi_i,p,p^{\prime}}
U_{p\phi_i}^{m}(U_{p^{\prime}\phi_i}^{m^{\prime}})^{\ast}e^{-\frac{i}{\hbar}%
(w_{\alpha}-w_{\alpha}^{\prime})t}\langle \mathbf{x}|w_{\alpha},m\rangle_{%
\phi_i}\langle w_{\alpha}^{\prime},m^{\prime}|\mathbf{x^{\prime}}%
\rangle_{\phi_i}
\end{equation}

\noindent where again as in the paper \cite{casati model} we have the
replacements $\mathbf{x}\longleftrightarrow \mathbf{x}-\frac{1}{2}\mathbf{s}$
and $\mathbf{x}\longleftrightarrow\mathbf{x^{\prime}}+\frac{1}{2}\mathbf{s}$
and also

\begin{equation}
\begin{split}
& \langle \mathbf{x}|w_{\alpha },m\rangle _{\phi _{i}}\sim e^{-\frac{i}{%
\hbar }(\mathbf{m}.\mathbf{x})} \\
& \langle w_{\alpha }^{\prime },m^{\prime }|\mathbf{x^{\prime }}\rangle
_{\phi _{i}}\sim e^{\frac{i}{\hbar }(\mathbf{m^{\prime }}.\mathbf{x^{\prime }%
})}
\end{split}
\label{amplitud cesaro prosen2}
\end{equation}%
\noindent From these two last equations \eqref{amplitud cesaro prosen1} and %
\eqref{amplitud cesaro prosen2} we have

\begin{equation}  \label{amplitud cesaro prosen3}
\langle \mathbf{x}|\widehat{\rho}(t)|\mathbf{x}\rangle=\sum_{i,p,p^{%
\prime},w_{\alpha},w_{\alpha}^{\prime}}\rho_{w_{\alpha},w_{\alpha}^{\prime},%
\phi_i,p,p^{\prime}}
U_{p\phi_i}^{m}(U_{p^{\prime}\phi_i}^{m^{\prime}})^{\ast} e^{-\frac{i}{\hbar}%
(\mathbf{m}-\mathbf{m^{\prime}}).\mathbf{x}}e^{\frac{i}{2\hbar}(\mathbf{m}+%
\mathbf{m^{\prime}}).\mathbf{s}}e^{-\frac{i}{\hbar}(w_{\alpha}-w_{\alpha}^{%
\prime})t}
\end{equation}

\noindent for the probability amplitude. In order to have an explanation in
terms of the ergodic level we must interpret this amplitude as the mean
value of some observable $\widehat{O}$, calculated in the state $\widehat{%
\rho }(t)$. If we choose $\widehat{O}=|\mathbf{x}\rangle \langle \mathbf{x}|$
then we have that

\begin{equation}  \label{amplitud cesaro prosen4}
\langle \mathbf{x}|\widehat{\rho}(t)|\mathbf{x}\rangle=\langle |\mathbf{x}%
\rangle\langle \mathbf{x}| \rangle_{\widehat{\rho}(t)}
\end{equation}

\noindent This means that the probability amplitude $\langle \mathbf{x}|%
\widehat{\rho }(t)|\mathbf{x}\rangle $ is the mean value of the projector $|%
\mathbf{x}\rangle \langle \mathbf{x}|$ in the state $\widehat{\rho }(t)$.
Now since the system is ergodic (eq. \eqref{Q-E}, \eqref{Q-E discrete}) we
expect that the time average of this amplitude is equal to the mean value of
$|\mathbf{x}\rangle \langle \mathbf{x}|$ in the state $\widehat{\rho }_{\ast
}$ in the limit $t\rightarrow \infty $. More precisely, if we use the
continuous version\footnote{%
Since in this case we have a continuous evolution given by the evolution
operator $\widehat{U}(t)$ we can choose any of the two versions of the
ergodic level (eq. \eqref{Q-E}, \eqref{Q-E discrete}). Both give the same
result.} (see \eqref{Q-E}), $\langle \mathbf{x}|\widehat{\rho }(t)|\mathbf{x}%
\rangle $ must satisfy

\begin{equation}
lim_{T\rightarrow \infty }\frac{1}{T}\int_{0}^{T}\langle \mathbf{x}|\widehat{%
\rho }(t)|\mathbf{x}\rangle dt=(\widehat{\rho }_{\ast }||\mathbf{x}\rangle
\langle \mathbf{x}|)=cte.  \label{amplitud cesaro prosen5}
\end{equation}%
\noindent This equation means that in the Casati-Prosen model the amplitude
tends in time average to the constant $(\widehat{\rho }_{\ast }||\mathbf{x}%
\rangle \langle \mathbf{x}|)$ and therefore that the average of the
interference term $P_{int}$ vanishes in time. Now we calculate the average
value of $\langle \mathbf{x}|\widehat{\rho }(t)|\mathbf{x}\rangle $. First,
from \eqref{amplitud cesaro prosen3} we can separate the sum of $%
\langle \mathbf{x}|\widehat{\rho }(t)|\mathbf{x}\rangle $ in a diagonal and
non-diagonal terms where the non-diagonal term is the interference term $%
P_{int}$. Then

\begin{equation}  \label{amplitud cesaro prosen6}
\langle \mathbf{x}|\widehat{\rho}(t)|\mathbf{x}\rangle=P_{diag}+P_{int}
\end{equation}

\noindent where

\begin{equation}  \label{amplitud cesaro prosen6}
\begin{split}
&P_{diag}=\sum_{i,p,w_{\alpha}}\rho_{w_{\alpha},\phi_i,p}|U_{p%
\phi_i}^{m}|^{2}e^{\frac{i}{\hbar}\mathbf{m}.\mathbf{s}}=cte. \\
&P_{int}=\sum_{i,p,p^{\prime},w_{\alpha}\neq
w_{\alpha}^{\prime}}\rho_{w_{\alpha},w_{\alpha}^{\prime},\phi_i,p,p^{%
\prime}} U_{p\phi_i}^{m}(U_{p^{\prime}\phi_i}^{m^{\prime}})^{\ast} e^{-\frac{%
i}{\hbar}(\mathbf{m}-\mathbf{m^{\prime}}).\mathbf{x}}e^{\frac{i}{2\hbar}(%
\mathbf{m}+\mathbf{m^{\prime}}).\mathbf{s}}e^{-\frac{i}{\hbar}%
(w_{\alpha}-w_{\alpha}^{\prime})t}
\end{split}%
\end{equation}

\noindent In the non-diagonal term $P_{int}$ of the equation
\eqref{amplitud
cesaro prosen6} we have assumed the crucial hypothesis of non-degeneracy $
w_{\alpha }\neq w_{\alpha }^{\prime }$ which is one of the mean features of the
chaotic billiards, e.g. the GOE, GUE and GSE spectral distributions (see for example \cite{stockmann}, \cite{casati libro}, \cite{haake}). The
non-degeneracy is a necessary condition for quantum chaos in the billiard
systems. Therefore, since the Casati-Prosen model is a chaotic billiard
then, in \eqref{amplitud cesaro prosen6} we cannot have terms of the type $%
\sum_{i,p,p^{\prime },w_{\alpha }}\rho _{w_{\alpha },\phi _{i},p,p^{\prime
}}U_{p\phi _{i}}^{m}(U_{p^{\prime }\phi _{i}}^{m^{\prime }})^{\ast }e^{-%
\frac{i}{\hbar }(\mathbf{m}-\mathbf{m^{\prime }}).\mathbf{x}}e^{\frac{i}{%
2\hbar }(\mathbf{m}+\mathbf{m^{\prime }}).\mathbf{s}}$. Then it is enough to
show that $lim_{T\rightarrow \infty }\frac{1}{T}\int_{0}^{T}P_{int}(\mathbf{x%
},t)dt=0$. We have

\begin{equation}
\begin{split}
&lim_{T\rightarrow \infty}\frac{1}{T}\int_0^{T}\langle\mathbf{x}|\widehat{%
\rho}(t)|\mathbf{x}\rangle dt=lim_{T\rightarrow \infty}\frac{1}{T}%
\int_0^{T}P_{diag}dt+ lim_{T\rightarrow \infty}\frac{1}{T}\int_0^{T}P_{int}(%
\mathbf{x},t)dt= \\
&=P_{diag}+lim_{T\rightarrow \infty}\frac{1}{T}\int_0^{T}P_{int}(\mathbf{x}%
,t)dt= \\
&=P_{diag}+ \\
&+\sum_{i,p,p^{\prime},w_{\alpha}\neq
w_{\alpha}^{\prime}}\rho_{w_{\alpha},w_{\alpha}^{\prime},\phi_i,p,p^{%
\prime}} U_{p\phi_i}^{m}(U_{p^{\prime}\phi_i}^{m^{\prime}})^{\ast} e^{-\frac{%
i}{\hbar}(\mathbf{m}-\mathbf{m^{\prime}}).\mathbf{x}}e^{\frac{i}{2\hbar}(%
\mathbf{m}+\mathbf{m^{\prime}}).\mathbf{s}}lim_{T\rightarrow \infty}\frac{1}{%
T}\int_0^{T}e^{-\frac{i}{\hbar}(w_{\alpha}-w_{\alpha}^{\prime})t}dt= \\
&=P_{diag}+ \\
&+\sum_{i,p,p^{\prime},w_{\alpha}\neq
w_{\alpha}^{\prime}}\rho_{w_{\alpha},w_{\alpha}^{\prime},\phi_i,p,p^{%
\prime}} U_{p\phi_i}^{m}(U_{p^{\prime}\phi_i}^{m^{\prime}})^{\ast} e^{-\frac{%
i}{\hbar}(\mathbf{m}-\mathbf{m^{\prime}}).\mathbf{x}}e^{\frac{i}{2\hbar}(%
\mathbf{m}+\mathbf{m^{\prime}}).\mathbf{s}} lim_{T\rightarrow \infty}\frac{%
i\hbar(e^{-\frac{i}{\hbar}(w_{\alpha}-w_{\alpha}^{\prime})T}-1)}{%
T(w_{\alpha}-w_{\alpha}^{\prime})} \\
&=P_{diag}=\sum_{i,p,w_{\alpha}}\rho_{w_{\alpha},\phi_i,p}|U_{p%
\phi_i}^{m}|^{2}e^{\frac{i}{\hbar}\mathbf{m}.\mathbf{s}}=(\widehat{\rho}%
_{\ast}||\mathbf{x}\rangle\langle\mathbf{x}|)=cte.
\end{split}%
\end{equation}

\noindent Then $lim_{T\rightarrow \infty}\frac{1}{T}\int_0^{T}P_{int}(%
\mathbf{x},t)dt=0$ and $\widehat{\rho}_{\ast}=\sum_{i,p,w_{\alpha}}\rho_{w_{%
\alpha},\phi_i,p} |U_{p\phi_i}^{m}|^{2}|w_{\alpha},m\rangle_{\phi_i}\langle
w_{\alpha},m|_{\phi_i}$.

We have obtained $\widehat{\rho }_{\ast }$ the Cèsaro limit of the initial
Gaussian wavepacket $\widehat{\rho }(0)$. Therefore we show that without
doing any kind of hypothesis about the energy spectrum, the equilibrium
limit of Casati-Prosen model is the Cèsaro limit $\widehat{\rho }_{\ast }$
and as a consequence the interference fringes vanish \textquotedblleft on
time average" for $t\rightarrow \infty $. This is physically expected since
the human eye averages in a scale time which is extremely larger than the
characteristic times of the decoherence of the quantum systems and therefore
we see that \textquotedblleft on time average" the interference fringes
vanish. This is the content of the ergodic level for the Casati-Prosen model
without any hypothesis about the energy spectrum.

Summarizing, the behavior of the Casati-Prosen model we can conceptually
explain, in the mentioned two ways, quasi-continuous approximation and
ergodic level. The first explanation is possible because its energy spectrum
that is discrete can be approximated by a quasi-continuous one. The two
explanations are not in contradictory but both differ in the type of
decoherence. In the quasi-continuous case it can be demonstrated that the
interference fringes vanish while using the ergodic level we strictly can
demonstrate that these vanish on time average. On the other hand, when the
quasi-continuous approximation is not valid then we can always use the
ergodic level. We remark the satisfying aspect is that both explanations are
based on the first two levels of the Quantum Ergodic Hierarchy, mixing and
ergodic respectively.

\subsection{The Kicked Rotator}

For the reasons that we have mentioned at the beginning the kicked rotator is one of most famous and studied chaotic systems.
The kicked rotator expresses in a simple way the physics of chaotic systems
whose Hamiltonian are of the type $H_{0}+H^{\prime }$ where $H_{0}$ is the
unperturbed and integrable Hamiltonian and $H^{\prime }$ is a delta time
periodic perturbation. This is important since the dynamics of many quantum systems
which presents chaos can be mapped into the kicked rotator. The Hamiltonian
is given by (see \cite{stockmann} eq. 4.2.1)

\begin{equation}  \label{kicked1}
H=\widehat{L}^{2}+\lambda cos\widehat{\theta}\sum_n \delta(t-n)
\end{equation}

\noindent which it describes the free rotation of a pendulum with angular
momentum $\widehat{L}$ periodically kicked by a gravitational potential of
strength $\lambda$. The moment of inertia $I$ and the kick period $T$ are
normalized to one. Classically, the kicked rotator presents different
behaviors for several values of the parameter $\lambda$. For small $\lambda$ values the
rotator shows regular behavior for most initial values of $\theta $ and $L$
with integrable regions of the space phase. But with increasing $\lambda$ the phase space becomes more and more chaotic, until for $\lambda>5$ most
regular parts have disappeared (see Fig. 1.3.(c) of \cite{stockmann}).

For a quantum mechanical description we need the evolution operator which is
given by the Floquet operator $\widehat{F}$ (see \cite{stockmann} eq. 4.2.12)

\begin{equation}  \label{kicked2}
\widehat{F}=e^{\frac{-i}{\hbar}\lambda cos\widehat{\theta}}e^{\frac{-i}{2\hbar}\tau
\widehat{L}^{2}}
\end{equation}

\noindent We note that since this evolution is discrete then we must use the
corresponding discrete versions of the levels of QEH. We can express any
initial state $\rho (0)$ in the eigenbasis of $\widehat{L}$, $|n\rangle =
\frac{1}{\sqrt{2\pi }}e^{in\theta }$

\begin{equation}
\rho (0)=\sum_{n}a_{n}|n\rangle \langle n|+\sum_{n\neq m}a_{nm}|n\rangle
\langle m|  \label{kicked3}
\end{equation}
In order to obtain the Césaro weak limit of $\widehat{\rho }(t)$ it is more
convenient to express the initial state of (106) in the eigenbasis $\left\{ |k\rangle \right\} $ of $\widehat{F}$.

\begin{equation}\label{kicked4}
\rho (0)=\sum_{k}\rho _{kk}|k\rangle \langle k|+\sum_{k\neq k^{\prime}}\rho _{kk^{\prime}}
|k\rangle \langle k^{\prime}|
\end{equation}

\noindent Now we show that $\widehat{\rho}_{\ast}=\sum_{k}\rho
_{kk}|k\rangle \langle k|$ is the Cèsaro limit of $\widehat{\rho}(t)$ for
all value of the parameter $\lambda$ and therefore, the kicked rotator belongs to
the ergodic level of the QEH for all $\lambda$.

$N$ successive applications of $\widehat{F}$ to $\widehat{\rho}(0)$ give the
state $\widehat{\rho}$ at the instant of time $t=N\tau$. More precisely, we
have

\begin{equation}  \label{kicked5}
\widehat{\rho}(N\tau)=\sum_k \rho_{kk} |k\rangle\langle k|+\sum_{k\neq
k^{\prime}}\rho_{kk^{\prime}}e^{-iN(\phi_k-\phi_{k^{\prime}})}|k\rangle%
\langle k^{\prime}|
\end{equation}

\noindent where the first and the second sums of \eqref{kicked5} are the
diagonal and non-diagonal terms of the state $\widehat{\rho }(N\tau )$, and
the phase $e^{-iN\phi _{k}}$ is the eigenvalue of the eigenstate $|k\rangle $
(see \cite{stockmann} page 136 and 137). Let $\widehat{O}$ be an observable.
From \eqref{kicked5} the mean value of $\widehat{O}$ in the state $\widehat{%
\rho }(N\tau )$ is

\begin{equation}  \label{kicked6}
\langle \widehat{O}\rangle_{\widehat{\rho}(N\tau)}=(\widehat{\rho}(N\tau)|%
\widehat{O})=Tr(\widehat{\rho}(N\tau)\widehat{O})=\sum_k \rho_{kk}
O_{kk}+\sum_{k\neq
k^{\prime}}\rho_{kk^{\prime}}e^{-iN(\phi_k-\phi_{k^{\prime}})}O_{kk^{\prime}}
\end{equation}

\noindent The second sum of \eqref{kicked6} are the interference terms, proper
of the quantum mechanical phenomena and the cancellation of this term is
an expression of \emph{decoherence}. From \eqref{kicked5} we have

\begin{equation}  \label{kicked7}
\begin{split}
&lim_{N\rightarrow\infty}\frac{1}{N}\sum_{j=0}^{N-1}(\widehat{\rho}(j\tau)|%
\widehat{O})=lim_{N\rightarrow\infty}\frac{1}{N}\sum_{j=0}^{N-1}\{\sum_k
\rho_{kk} O_{kk}+\sum_{k\neq
k^{\prime}}\rho_{kk^{\prime}}e^{-ij(\phi_k-\phi_{k^{\prime}})}O_{kk^{%
\prime}}\}= \\
&=\sum_k \rho_{kk} O_{kk}+\sum_{k\neq
k^{\prime}}\rho_{kk^{\prime}}O_{kk^{\prime}}lim_{N\rightarrow\infty}\frac{1}{%
N}\{\sum_{j=0}^{N-1}(e^{-i(\phi_k-\phi_{k^{\prime}})})^{j}\}= \\
&=\sum_k \rho_{kk} O_{kk}+\sum_{k\neq
k^{\prime}}\rho_{kk^{\prime}}O_{kk^{\prime}}lim_{N\rightarrow\infty}\frac{%
1-e^{-iN(\phi_k-\phi_{k^{\prime}})}}{N(1-e^{-i(\phi_k-\phi_{k^{\prime}})})}%
=\sum_k \rho_{kk} O_{kk}=(\widehat{\rho}_{\ast}|\widehat{O})
\end{split}
\end{equation}

\noindent where we have used that $lim_{N\rightarrow\infty}\frac{%
1-e^{-iN(\phi_k-\phi_{k^{\prime}})}}{N(1-e^{-i(\phi_k-\phi_{k^{\prime}})})}%
=0 $ for all $k,k^{\prime}$ and that $\widehat{\rho}_{\ast}=\sum_k \rho_{kk}
|k\rangle\langle k|$. From \eqref{kicked7} and the discrete version of the
ergodic level \eqref{Q-E discrete} the kicked rotator is
ergodic for all $\lambda$ and the Cèsaro limit is $\widehat{\rho}_{\ast}=\sum_k \rho_{kk}
|k\rangle\langle k|$ which is the equilibrium state ``on time-average". In
this time-average sense we can say that the kicked rotator decoheres to the
state $\sum_k \rho_{kk} |k\rangle\langle k|$ for all initial state and the
decoherence time is $\infty$.

Now we see that for values of $\lambda>5$ the behavior is different from the
ergodic case. In such case the expected quantum mechanically distribution $
f_{N}(L)$ for the quadratic mean value of the angular momentum $\langle
\widehat{L}^{2}\rangle $ after $N$ kicks is given by (see \cite{stockmann}
eq. 4.2.20)

\begin{equation}  \label{kicked8}
f_N(L)=\frac{1}{l_s}e^{-\frac{2|L|}{l_s}}
\end{equation}

\noindent This exponential localization implies that for kick numbers $N\leq
l_{s}$ we are in the range of classical diffusion and for $N\gg l_{s}$ we
are in the fully chaotic behavior \footnote{In the classical sense. In fact, when $\lambda>5$ the phase space of its classical analogue is chaotic with some surviving stable islands (see \cite{stockmann} pag. 10).}. For $N\gg l_{s}$ the phase factors
$e^{-iN(\phi _{k}-\phi _{k^{\prime }})}$ of \eqref{kicked6} oscillate rapidly
in such a way that only survive the terms with $k=k^{\prime }$, that is

\begin{equation}  \label{kicked9}
\langle \widehat{O}\rangle_{\widehat{\rho}(N\tau)}=(\widehat{\rho}(N\tau)|%
\widehat{O})\simeq \sum_k \rho_{kk} O_{kk}=(\widehat{\rho}_{\ast}|\widehat{O}%
) \,\,\,\ for \,\ N\gg l_s
\end{equation}

\noindent where $\widehat{\rho}_{\ast}=\sum_k \rho_{kk} |k\rangle\langle k|$. Then from \eqref{kicked9} in the case $\lambda>5$ we have

\begin{equation}  \label{kicked10}
lim_{N\rightarrow\infty}(\widehat{\rho}(N)|\widehat{O})=lim_{N\rightarrow%
\infty}(\widehat{\rho}(N\tau)|\widehat{O})=\sum_k \rho_{kk} O_{kk}=(\widehat{%
\rho}_{\ast}|\widehat{O})
\end{equation}

\noindent Now by the mixing condition (eq.\eqref{quantum mixing condition
discrete}) for the discrete case, the equation \eqref{kicked10} says that
for $\lambda>5$ the kicked rotator belongs to the mixing level and the equilibrium
state is the weak limit $\widehat{\rho}_{\ast}=\sum_k \rho_{kk}
|k\rangle\langle k|$. $\widehat{\rho}_{\ast}$ is the diagonal part of the
initial state $\widehat{\rho}(0)$ (see eq. \eqref{kicked4}) written in the
basis of the Floquet operator $\widehat{F}$. This situation is analogue to
SID \cite{SID} where the decoherence is performed in the energy basis.
Meanwhile, for the fully chaotic regime \footnote{Again, in the classical sense.} $\lambda>5$ the kicked rotator decoheres
in the Floquet basis $\{|k\rangle\}$.

Moreover, from equation \eqref{kicked9} and the Bernoulli condition %
\eqref{Bernoulli condition} we have $(\widehat{\rho }(t)|\widehat{O}%
)=\sum_{k}\rho _{kk}O_{kk}$ \,\,\,\,\,\,\,\,\,\,\,\,\,\,\,\,\,\,\ $=(\widehat{\rho }_{\ast }|\widehat{O})$ for $%
t=N\tau \gg \tau l_{s}$. That is, the kicked rotator is Bernoulli from any
time $t\gg \tau l_{s}$. And this characteristic time $t_{D}=\tau l_{s}$ expresses the decoherence of the kicked rotator. In this way we
conclude the kicked rotator analysis in terms of the Quantum Ergodic Hierarchy.

\section{Conclusions}

\label{s:Conclusions}
In this paper we have introduced a definition of the four main levels of the
Quantum Ergodic Hierarchy with the property that their classical limits are
the corresponding usual levels of the classical ergodic hierarchy.

Language translation of the sigma algebras of the Kolmogorov level to
quantum language could be made and reduced to a single condition (see
equation \eqref{quantum Kolmogorov condition}) thanks to the application of
the principle of inclusion-exclusion (see equation
\eqref{exclusion-inclusion}), which helped us to extend the technique used in
the paper \cite{0} for the ergodic level and mixing level. Here we have used
the properties of the Wigner transform. Thus the resulting condition for the
Kolmogorov quantum level is consistent with the definitions of mixing and
Bernoulli (see 5.2 and 6.2 sections).

Language translation of the Bernoulli level was the most immediate of all
levels of the hierarchy ergodic. However, additionally we have translated
the \emph{independence events property} of Bernoulli systems (see equation
\eqref{independence}) into a quantum version in the sense of the expectation
values (see equation \eqref{bernoulli4}). The physical interpretation of this factorization is the quantum no-correlation between the observables of a product of the type \eqref{bernoulli4}. This property was necessary to demonstrate the
inclusion of quantum Bernoulli level within the quantum Kolmogorov level
(see section 6.2).

We have just translated the four levels of the ergodic
hierarchy to quantum language, and these levels from the lowest (ergodic) to the highest (Bernoulli) are schematized in the following diagram where the inclusions are strict.

\vskip0.7truecm

\begin{equation}
ERGODIC\supset MIXING\supset KOLMOGOROV \supset BERNOULLI
\end{equation}

\vskip0.7truecm

\noindent In the next table we list the
Quantum Ergodic Hierarchy levels in a compact way.

\vskip1truecm

\centerline{TABLE II: \,\ THE QUANTUM ERGODIC
HIERARCHY (QEH)}\vskip1truecm
\begin{tabbing}
\hspace*{3cm} \= \hspace*{8cm} \= \hspace*{4cm} \kill
\ \textsc{LEVEL} \> CONDITION EQUATION \> PROPERTIES \\\\
\textbf{\emph{Ergodic}} \> $\lim_{T\rightarrow \infty }\frac{1}{T}\int_{0}^{T}(\widehat{\rho }(t)|\widehat{O})dt=(\widehat{\rho }_{\ast }|\widehat{O})$ (continuous) \> Cesaro limit equals to $\widehat{\rho}_{\ast}$ \\\\
      \> $\lim_{N\rightarrow \infty }\frac{1}{N}\sum_{k=0}^{N-1}(\widehat{\rho }(k)|\widehat{O})dt=(\widehat{\rho }_{\ast }|\widehat{O})$ (discrete)\> \\\\
\textbf{\emph{Mixing}} \> $\lim_{t\rightarrow \infty }(\widehat{\rho}(t)|\widehat{O})=(\widehat{\rho }_{\ast }|\widehat{O})$ (continuous)\> Weak limit equals to $\widehat{\rho }_{\ast}$\\\\
      \> $\lim_{N\rightarrow \infty }(\widehat{\rho}(N)|\widehat{O})=(\widehat{\rho }_{\ast }|\widehat{O})$ (discrete)\> \\\\
\textbf{\emph{Kolmogorov}} \> $\lim_{n\rightarrow\infty}\{(\widehat{\rho}(n+m_1)|\widehat{O}_1\prod_{j=2}^{\infty}\widehat{O}_j(n+m_j))$\>  Weak limit equals to $\widehat{\rho }_{\ast}$\\\\
      \> $-\prod_{j=2}^{\infty}(\widehat{\rho}(n+m_1)|\widehat{O}_j(n+m_j))(\widehat{\rho_{\ast}}|\widehat{O}_1)\}=0$ \> \\\\
\textbf{\emph{Bernoulli}} \> $(\widehat{\rho}(t)|\widehat{O})=(\widehat{\rho }_{\ast}|\widehat{O})$ \> $(\widehat{\rho}(t)|\prod_{i}\widehat{g_i})=\prod_{i}(\widehat{\rho}(t)|\widehat{g_i})$\\
\end{tabbing}

\noindent In this table we can see the level of complexity of the condition
that defines each level of the ergodic hierarchy. Starting at the lowest
level, the ergodic, which translates into an average temporal of expectation
values following by the mixing level corresponding to weak limit. And
continuing with Kolmogorov level that represents a condition on a set of
observables (the language translation of the sigma algebra) and ending with
the Bernoulli level representing the null correlation for all time.

In section 7 we have presented two emblematic examples that show the relevance of the
Quantum Ergodic Hierarchy: the Casati-Prosen model and the kicked rotator.
We have explained their chaotic and decoherence behavior in terms of the QEH
levels in a conceptual way.

For the Casati-Prosen model, the cancellation of the interference fringes
can be deduced by the chaotic nature of its classical analogue in two ways,
mixing and ergodic respectively: In the first case, for large times $t$ but shorter than
$t^*$ (see section 7.1) the spectrum can be approximated by a quasi-continuous one and in the
second case, otherwise we can maintain the spectrum discrete. But in both cases the equilibrium
state $\widehat{\rho }_{\ast }$ expresses the type of decoherence of the
initial Gaussian wavepacket $\widehat{\rho }(0)$, weak limit in the first
case and Cèsaro limit in the second case.

For the kicked rotator we have characterized its quantum chaos transition,
from small values of $\lambda$ to greater than 5, in terms of the Quantum Ergodic
Hierarchy levels corresponding to each regime. This characterization is
summarized in the next scheme.

\begin{equation}
\begin{split}
&\,\,\,\,\,\,\,\,\,\,\,\,\,\,\,\,\,\,\,\,\,\,\,\,\,\,\,\,\,\,\ \lambda\ll1 \,\,\
\Longrightarrow \,\,\ integrable, \,\ regular \,\ behavior \,\,\
\Longrightarrow \,\,\ \mathbf{ergodic} \,\ level \\
& \\
&\lambda\sim \lambda_c=0,9716... \,\,\ \Longrightarrow \,\,\ stochastic \,\ and \,\
diffusive \,\ behavior \,\,\ \Longrightarrow \,\,\ \mathbf{ergodic} \,\ level
\\
& \\
&\lambda>5\,\,\ \Longrightarrow \,\,\ fully \,\ chaotic, \,\ exponential \,\
localization \,\,\ \Longrightarrow \,\,\ \mathbf{mixing} \,\ and \,\ \mathbf{%
Bernoulli} \,\ levels
\end{split}%
\end{equation}

\noindent We see that for the first two regimes, small values of $\lambda$ and
values of $\lambda$ near to the critical value $\lambda_{c}=0,9716$ (see \cite{stockmann}
page. 10 and 145), the regular and stochastic-diffusive behavior both
correspond to the ergodic level of the QEH. Then QEH can not differentiate
this two regimes. This is so because the Quantum Ergodic Hierarchy is a
chaos classification for large times (in the limit $t\rightarrow \infty $)
and therefore, we can not expect the chaos effects would occur in a finite
interval of time.

Moreover, QEH allows to classify chaotic phenomena of large characteristic
times as the decoherence of the Casati-Prosen model and the exponential
localization of the kicked rotator for times $t\gg \tau l_{s}$. The
exponential localization of the fully chaotic regime ($\lambda>5$ and $t\gg \tau
l_{s}$) of the kicked rotator corresponds to the Bernoulli level which is the most chaotic. On the positive side of
The Quantum Ergodic Hierarchy we see that the regime less chaotic ($%
\lambda\lesssim \lambda_{c}$) and more chaotic ($\lambda>5$) correspond to the lowest and
highest levels of QEH, ergodic and Bernoulli respectively. We consider that
this agreement of QEH with other approximations to quantum chaos like the
computational complexity in the Casati-Prosen model or the Floquet Theory in
the kicked rotator are a positive first step of QEH as an alternative
theoretical framework to study the phenomena of the quantum chaos.

\vskip1truecm

\noindent \textbf{Acknowledgements} \noindent This paper was partially
supported by the CONICET (Argentine Research Council) the FONCYT (the
Argentine Fond for the Research) and the Buenos Aires University.

\end{document}